# Cold free radical molecules in the laboratory frame

J. R. Bochinski, Eric R. Hudson, H. J. Lewandowski, and Jun Ye

JILA, National Institute of Standards and Technology and Department of Physics, University of Colorado, Boulder, CO 80309-0440

A special class of molecules that are important to many subfields in molecular dynamics and chemical physics, namely free radical molecules, now enjoy a significant degree of center-of-mass motion control in the laboratory frame. The example reported in this paper concerns hydroxyl radical (OH), which, after the internal degrees of freedom are cooled in a supersonic expansion, has been bunched, accelerated, and slowed using time-varying inhomogeneous electric fields. *In situ* observations of laser-induced fluorescence along the beam propagation path allows for detailed characterization of the longitudinal phase-space manipulation of OH molecules by the electric fields. The creation of a pulse containing 10^3 – 10^6 molecules possessing a longitudinal velocity spread from 2 to 80 m/s around a mean laboratory velocity variable from 550 m/s to rest with only a few mm spatial extent represents an exciting and useful new experimental capability for exploring free radical dynamics. This paper offers the most detailed study of the Stark deceleration dynamics to date.

## Introduction

Efforts to create, study, and utilize samples of cold molecules are at the forefront of experimental atomic and molecular physics[1 2 3 4 5]. Scientific explorations of cold molecular samples will allow spectroscopy to be performed at the fundamental level where vibrational relaxation can be directly revealed and extremely weak interactions can be investigated. Metrological applications will surely follow. The long interrogation times that cold molecules will provide is critical to a number of precision measurement projects such as the search for the permanent electron electric dipole moment[6 7] and improved values of fundamental constants like the fine structure $\alpha$. At low translational energies, molecular orientation will play an important role in novel cold collisions[8 9 10 11 12] and ultracold molecular dynamics[13]. Observations of cold atom–molecule interactions may also reveal new regimes in physical chemistry[14]. Further, there has been a proposed application in quantum computing[15] for trapped ultracold polar molecules.

Heteronuclear molecules possess permanent electric dipole moments that create new opportunities for control of intermolecular interactions. While a number of research groups are actively pursuing experiments to form ultracold polar molecules utilizing photoassociation techniques to create weakly bound, vibrationally excited molecules from double-species magneto-optical traps of cold atoms[16 17 18], some other groups have focused on ground-state molecules that are manipulated by buffer gas cooling[1 19 20] or Stark deceleration[21 22 23 24 25]. In this paper we focus on a particular species of stable molecules, namely the diatomic hydroxyl radical (OH). We note, as with most molecular systems, it is not technically practical to create cold OH samples using the abovementioned cold atom approach. Nevertheless, while alternate techniques must be employed to produce the cooled sample, neutral free radicals remain an extremely attractive and significant class of particles for cold molecule research. Historically important as a focal point of study in physical chemistry[26 27], astrophysics[28 29], atmospheric[30] and combustion physics[31 32 33], these chemically reactive molecules typically possess large electric and magnetic dipole moments, thus supporting a variety of interactions. In the low temperature regime, these interactions may create interesting effects; for example, at temperatures of ~1 μK, the translational energy becomes less than the intermolecular electric dipole-dipole interaction energy. Considering that the relative orientation of these dipoles determines whether the interaction is attractive or repulsive, these polar molecules can easily influence each other's trajectory and lead to complex dynamics. In principle, an external electric field can then be used to guide collisions and perhaps even to exert a new means for management of chemical reactions. Interestingly, this mechanism represents a profoundly different type of control over collisions than the usual resonant scattering between cold atoms exerted by a magnetic field. The relatively strong dipolar interaction should also open new analogies to condensed matter systems, particularly those systems that interact by strong spin-dependent forces[34]. It may even be possible that the cold molecules could form a "dipolar crystal", in analogy with an ionic crystal formed in ion traps[35]. Finally, we note experimental conditions under which both electric and magnetic fields simultaneously interact with the cold molecules may naturally provide even more useful and interesting dynamics to investigate[36].

Hence, production of a large collection of neutral free radical molecules populating a single quantum state and having a small center-of-mass motion in the laboratory frame represents a critical first step towards realizing these potential cold molecule experiments. We employ the technique of Stark deceleration to prepare cold packets of the OH molecules. Molecules produced at low temperatures by the Stark slowing process will be in their rovibrational ground states, and therefore do not suffer from rovibration-changing collision losses. However, at low temperature limit and under certain external electric fields and sufficient densities, the molecules in a weak-field seeking state can still experience non-negligible inelastic scattering losses to strong-field seeking states [9]. The Stark deceleration approach[21 22 23 24 25] has worked effectively with molecules such as CO, $ND_3$, OH, and YbF, resulting in experimental observations of slowing and trapping. Our previous work[24] demonstrated deceleration of stable packets of OH, with a minimum translational velocity in the laboratory frame of 140 m/s and a velocity spread less than 2.7 m/s. In the present work, we report further progress along this research direction, culminating with the present capability to produce phase-stable packets of molecules with a molecular-frame temperature from ~1K to ~ tens of millikelvin around a mean laboratory velocity variable from 550 m/s to essentially zero. We also document detailed technical approaches at various stages of our cold molecule machine.

The complex structure of the molecular states and the associated energy levels of OH will be described in the next section. The Stark shift of the relevant molecular states due to an external electric field is also introduced, along with the molecular detection scheme based on sensitive, laser-induced fluorescence (LIF) measurement. Section III gives a general description of the entire experimental setup, with a brief discussion regarding construction of the supersonically cooled OH beam source, which combines a novel discharge apparatus with a pulsed supersonic expansion valve. Armed with the understanding of the Stark effect of OH, we describe in Section IV beam propagation into the acceptance aperture of the decelerator. This process is aided by an electric hexapole–based molecular lens that provides initial state selection and focusing. Section V provides underlining principles and mechanisms responsible for the Stark deceleration process. Theoretical descriptions of the longitudinal phase-space evolution and comprehensive numerical simulations using realistic molecular parameters and modeled spatial distributions of inhomogeneous electric fields are presented. Practical information concerning the design, fabrication, construction, and preparation of the Stark decelerator is given in Section VI. The simulation model presented in Section V generates results that provide excellent agreement to the experimental observations reported in Section VII. The *in situ* LIF detection enables detailed characterization of the experimental phase-space evolution of the molecular packets at various stages along the entire decelerator and within the quadrupole electrostatic trap region. These observations demonstrate a thorough understanding and control of the molecular slowing process and reveal that the present experimental system is fully functional for ongoing research in molecular trapping and cold molecule collisions. Finally, in Section VIII we discuss future directions of the experiment involving improved electrostatic and magnetic trapping, limited laser cooling, as well as collision experiments between trapped atomic and molecular samples.

## Structure of OH

The neutral hydroxyl free radical (OH), though extremely chemically reactive, is a stable, diatomic molecule. As the basic molecular energy structure is well understood, here we will discuss only the states relevant to the current experiment. For low rotation levels, the angular momentum coupling of the $^2\Pi$ electronic ground state of the OH molecule is sufficiently described by Hund's case (a)[37]. Under this interaction scheme, both the electron orbital angular momentum **L** and electron spin **S** are coupled to the intramolecular axis, leading to the definition of the total electron angular momentum as $\mathbf{\Omega} = \Lambda + \Sigma$, where $|\Lambda| = 1$ ($|\Sigma| = 1/2$) represents the projection of **L** (**S**) onto this axis. This coupling results in two spin-orbit states, where the $^2\Pi_{3/2}$ state lies ~126 cm$^{-1}$ below the $^2\Pi_{1/2}$ state as shown in Figure 1. Strong spin-orbit coupling ($A_{SO}$ = -139 cm$^{-1}$) is responsible for the observed large splitting between the two $\Omega$ branches. The molecular angular momentum **J** in the laboratory frame is defined as $\mathbf{J} = \mathbf{L} + \mathbf{S} + \mathbf{R}$, where **R** represents the nuclear rotation angular momentum. The allowed values of **J** in the laboratory frame are given as $J = \Omega, \Omega + 1, \Omega + 2, \ldots$ The relatively large energy splitting between **J** states is due to the small moment of inertia of the molecule and subsequent large rotational constant. The coupling between the nuclear rotation and the electron angular momentum (projection of **L** along the internuclear axis) results in lambda-type splitting ($\lambda$-doubling) of each state, denoted as f and e states. In the figure, these energy separations are exaggerated for clarity. For $^{16}$OH with an intrinsic nuclear spin **I** of 1/2, the free radical molecule is a boson. The total angular momentum is given by $\mathbf{F} = \mathbf{I} + \mathbf{J}$. Each $\lambda$-doublet component is split into hyperfine states that are characterized by a symmetry index p that defines the state's parity (±). The hyperfine energy separations have also been exaggerated for clarity of presentation.

The $^2\Sigma^+$ first electronic excited state of the OH molecule has an electron spin of 1/2 but no electron angular momentum; thus, the appropriate couplings are well-described by Hund's case (b). For this angular momentum configuration, the only important interaction is between the nuclear rotation of the molecule **R** and the electron spin **S** -- so-called $\rho$-doubling. This coupling separates every **J** state that has non-zero **R** into two states of the same parity. Figure 1 shows the important energy levels and schematically depicts the excitation (solid arrow) and detection (dashed arrow) pathways, as described in detail below.

Because the $\lambda$-doublet states possess opposite parity and small energy separations, adjacent levels of the ground state OH readily mix under the interaction of the molecule's permanent electric dipole (1.67 D[38]) with an external electric field. The energy of these mixed states increases (weak-field seeking states) or decreases (strong-field seeking states) as a function of the electric field strength, as shown for the rovibrational ground state of OH in Figure 2(a). The Stark energy shift evolves from quadratic to linear when the energy change due to the applied field exceeds the λ-doublet splitting value of 0.055 $cm^{-1}$. The ground state shift is unaffected by the next higher rotational state due to the large energy separation of ~ 84 $cm^{-1}$. In the high electric field regime, the complex energy levels depicted in Fig. 2(a) emerge as four distinct families. As shown in Fig. 2(b), for the Stark slowing experiment discussed here, the weak-field seeking states are most relevant, with the $^2\Pi_{3/2}$ F = 2, $|m_F|$ = 2,1 states experiencing approximately three times the Stark energy shift of the $^2\Pi_{3/2}$ F = 2, $|m_F|$ = 0 and $^2\Pi_{3/2}$ F = 1, $|m_F|$ = 0,1 states. The horizontal dashed line corresponds to the zero energy level while the vertical dotted line represents the peak electric field magnitude under the normal operating voltages of the Stark decelerator. We note molecules in strong-field seeking states do not survive the first state-selection device, namely the hexapole acting as a molecular focusing lens, and therefore do not contribute to any observed signals thereafter.

Optical detection via fluorescence measurement is a powerful tool in atomic and molecular physics experiments. A relatively favorable branching ratio, a large difference between the excitation and decay wavelengths, and a reasonable quantum efficiency in fluorescence detection make laser induced fluorescence (LIF) a versatile approach for measuring the hydroxyl radical. The excitation source is tuned resonant with the $X^2\Pi_{3/2}$ (v = 0) → $A^2\Sigma^+$ (v = 1) electronic transition at 282 nm for subsequent detection of the red-shifted radiative decay around 313 nm $A^2\Sigma^+$ (v =1) → $X^2\Pi_{3/2}$ (v = 1), as shown in Figure 1. The separation in wavelength allows effective optical filtering to reduce the background scattering, as discussed in detail in the next section. The characteristic fluorescence 750 ns decay lifetime provides a distinctive signature of the presence of OH molecules.

## OH beam source and detection

Supersonic expansion in molecular beam experiments is a widely used technique. Under proper operating conditions, rotational and vibrational temperatures are significantly lowered along with the benefit of a reduced translational velocity spread in the molecular frame. Within the present experiment, collisions during the expansion represent the only true cooling mechanism. The maximum phase-space density achievable in the experiment is determined at this stage since during the subsequent deceleration process the phase-space distribution of the molecules undergoes conservative rotation without any enhancement in density. Therefore, as long as the relatively large translational speed of the molecular beam can be compensated for by the slowing capability of the Stark decelerator, a supersonic expansion provides a very useful initial source for creation and experimentation of cold molecules.

Figure 3 illustrates the overall experimental setup, showing the discharge and the pulsed valve configuration, along with the subsequent skimmer, hexapole, decelerator stages, and the electrostatic quadrupole trap. The experiment utilizes a seeded, pulsed supersonic beam combined with electric discharge to realize an efficient free radical source. The source apparatus creates an intense OH molecular beam, possessing both narrow initial temporal width and small longitudinal velocity spread, primarily populating only the lowest rotational ground state energy level. Xenon (Xe), due to its large mass, is employed as an inert carrier gas to minimize the initial molecular velocity. At a stagnation pressure of ~3 atmospheres, Xe bubbles through a small reservoir of deionized (> 10 MΩ cm), distilled water ($H_2O$). When the valve opens, the carrier gas (~99%) and seeded $H_2O$ (~1%) expand supersonically into the vacuum chamber. The 0.5 mm diameter exit nozzle valve is operated at a 5 Hz repetition rate, creating a molecular supersonic beam with a full pulse width of <100 μs.

We utilize an electric discharge to generate the hydroxyl radical molecules *during* the supersonic expansion. Such methods have been used previously[39 40 41]. Our approach[42] contains several key ingredients, including the use of a hot filament to aid in discharge initiation, selected voltage polarity for maximum discharge stability, geometry of the discharge electrodes to ensure *where* the discharge creates molecules is well-defined, as well as utilization of a μs-scale high voltage (HV) pulse to control precisely *when* the discharge occurs. These beneficial improvements over earlier work lead to a dramatic reduction in the discharge heating of the molecules, resulting in the production of a rotationally and longitudinally colder radical beam, well-suited for the deceleration experiment.

For the most stable discharge operation, the voltage polarity is selected such that electrons are accelerated against the propagation direction of the expanding supersonic jet. Application of sufficiently high voltage to the electrodes during the molecular supersonic expansion enables self-initiating of an electric discharge that produces the hydroxyl radical (OH) by disassociating the $H_2O$[43]. When the OH creation occurs spatially near to the nozzle in the expansion region, the diatomic molecule is subsequently cooled by collisions with the more abundant Xe atoms, resulting in rotationally cold free radicals. Empirically, we learned that operating the discharge voltage in a short duration, pulsed manner significantly reduces undesirable heating of the molecular beam by the discharge. Moreover, employing a current-driven filament within the vacuum chamber in addition to the pulsed high voltage ignites the discharge at a significantly lower voltage level, further minimizing heating to molecules during the supersonic expansion. We utilize a fast switch to pulse the high voltage (typically, for 2 μs at -1.4 kV), resulting in increased OH production, lower rotational temperature, narrower longitudinal velocity spread, and reduced mean longitudinal velocity. Under normal operating conditions, the source produces a free radical molecular pulse having a mean speed around 385 m/s with a FWHM velocity spread of 16% at a rotational temperature of < 27 K, corresponding to ~ 98.5% of the molecules in the J = 3/2 lowest rotational state.

The filament–assisted discharge can occur at virtually any molecular density during the supersonic expansion. Hence, the narrow-pulsed HV discharge can selectively create OH over a broad range in time within the ~100 μs wide molecular pulse envelope. As the discharge duration is much shorter than the molecular pulse, OH molecules generated at different times within the pulse reflect the different instantaneous characteristics of the supersonic expansion at the moment of their creation. This effect is most directly manifested as the ability to tune the mean longitudinal velocity of the molecular pulse, as depicted in Figure 4. In this figure, pulses of rotationally cold OH are measured by LIF after they have propagated from the source to just in front of the molecular beam skimmer. The pulses were created using identical discharge voltage amplitudes and pulse widths but varying the discharge initiation time within the molecular pulse. This initial creation time difference has been normalized for the three data traces where time zero represents the discharge initiation; the subsequent change in arrival time demonstrates the differences in pulse mean velocity. As the discharge initiation time is varied from earlier to later within the overall molecular pulse envelope, the amplitude and mean speed of the OH signal decreases while the velocity width increases, consistent with sampling OH production within the diminishing supersonic expansion.

Potentially, background-free resonance fluorescence is a highly sensitive technique for detecting small numbers of molecules. In analytical chemistry and biology, resonance fluorescence is used extensively for efficient detection and identification of single molecules[44]. The sensitivity of the fluorescence detection technique depends upon the characteristics of the decay channels of the excited states, namely that the radiative branching ratios are favorable and quenching processes are not significant. Further, for maximum detected signal, the decay photons should be collected over as large a solid angle as possible; in our experiment, intra-vacuum optics provide solid angle efficiencies of 0.5-4.5%, depending on the specific spatial location. With quantum efficiencies in the range of 10% for the cathode surfaces of modern photomultipliers, one then expects a photoelectron event for every ~200 fluorescent decays. The dark current of the quantum detector is a few to a few tens of counts per second. However, various phenomena may degrade the signal-to-noise ratio. Even with careful baffling and shielding, the most severe limitation on the detection sensitivity in the present experiment is due to scattered excitation laser photons, which dominate stray background light, even after careful imaging.

A key benefit of utilizing fluorescence detection as opposed to a spatially fixed ionization-based scheme is the ability to detect *in situ* molecules at multiple locations, greatly increasing the experimental flexibility and level of understanding. OH signals are measured *within* the decelerator *during* the actual slowing process, giving insight into the active longitudinal phase-space manipulation of the molecules by the pulsed, inhomogeneous electric fields. Specifically, in the experiment we employ LIF detection to monitor the hydroxyl radicals after the discharge, observe the focusing effects of the hexapole on the molecules, measure the molecular packet spectra at various positions within the Stark decelerator, and finally, detect the molecules within the electrostatic trap region.

Using a beta barium borate (BBO) crystal to frequency-double a 564 nm dye laser output (~10 ns pulse duration) creates the ultra violet (UV) excitation light at 282 nm with high peak intensity. The excitation laser is counter-propagated to the molecular pulse traveling down the central axis of the Stark decelerator. The relatively spectrally broad light source (~2 GHz) encompasses any Doppler-induced frequency shifts in the OH resonance as the molecular longitudinal velocity changes during the deceleration process. Reduction of the laser scatter is paramount for optimized signal collection. First, due to the separation of the excitation and fluorescence wavelengths, the carefully selected photomultiplier tube (PMT) has a photocathode responsivity that is severely reduced at the laser wavelength versus the fluorescence wavelength. Next, an interference filter selectively inhibits the laser light ($10^3$ suppression) versus the fluorescent photons (~ 20% transmission coefficient). We also utilize a *switched* PMT whose dynode voltages are quickly arranged during the laser pulse so as to actively repel photoelectrons liberated by the scattered UV light on the cathode material. The first, third, and fifth dynodes are used rather than the photocathode itself for improved switching speed, while the rest of the elements in the dynode chain remain at their normal voltages. The usual state of the PMT is "interrupted"; in order to observe the OH molecular decay signal an externally applied control pulse rapidly turns the detector "on" for ~5 μs with the appropriate dynode voltage levels for normal PMT operation. In this manner, a ~$10^4$ suppression of the measured laser intensity is achieved, at a small expense of a fast noise spike generated by the switching currents. This repeatable noise feature caused by the PMT switching is subtracted off as a background signal. We note additionally, this system also works well using fewer switched dynodes. For example, utilizing only two dynodes (e.g., the first and fourth dynode) provides attenuation of laser scatter by ~$10^3$.

The OH molecular signal is measured as a function of time at a particular spatial location in the following manner. First, immediately prior to the measurement (~ 1 μs), all high voltage electrodes within the vacuum chamber are grounded to avoid Stark shifting of the OH transition frequency. Subsequently, before the molecules have had an opportunity to exit the specific detection region, the excitation laser pulse fires, generating the OH fluorescence signal. The detected photoelectrons from the PMT are amplified, averaged 300 times, and displayed on a digital oscilloscope over a ~5 μs time window to fully encompass the appropriate state decay lifetime. Next, a background spectrum consisting of an equal number of averaged traces with the source discharge voltage off (and thereby, consisting of laser scatter, switching, and other intrinsic noise contributions) is subtracted from the measured signal, resulting in the corrected molecular fluorescence spectrum. This spectrum is then integrated by the oscilloscope, giving a value directly proportional to the number of OH molecules present in the detection area at that time. The laser is stepped later in time relative to when the valve opens, and the entire measurement is repeated. In this iterative manner, a spectrum consisting of the OH signal as a function of time is generated.

Figure 5 shows time-of-flight (TOF) LIF measurements of OH molecules at different spatial locations in the experiment as indicated in Figure 3. Figure 5 trace (a) shows OH signals detected just before the skimmer (1.58 cm in height, 1.5 mm diameter central aperture) tip. The peak amplitude of Fig. 5(a) corresponds ~3 x $10^6$ molecules at a density of ~$10^9$ $cm^{-3}$, with an total pulse population ~3 x $10^7$ molecules. With no voltages present on the molecular focuser, the OH packet traverses the skimmer and hexapole region as shown in Figure 5 trace (b). The approximate two orders of magnitude signal loss is due to transverse spreading of the OH beam. Operating the pulsed hexapole with high voltages provides state selection and transverse focusing, leading to enhanced signal sizes after the hexapole. Measurements are made further downstream when the hexapole is in operation, as well as with constant voltages applied to all electrode stages. These unswitched voltages result in no net longitudinal velocity change but provide transverse confinement to the molecular pulse as it propagates down the slower; we refer to this mode of operation as "transverse guidance." Figure 5, traces (c), (d), and (e) are observed OH signals as the pulse propagates before the 23$^{rd}$ stage, the 51$^{st}$ stage, and into the center of the electrostatic trap region. The increase in molecular pulse width corresponds to the expected "stretching" in the longitudinal direction that occurs under free flight as faster molecules move ahead and slower molecules lag behind. Integration of measured peak areas at different observation points reveal transverse loss of molecules within the slower, as discussed in more detail in Section VII (e.g., see Figure 10(d)).

## Electric hexapole as a molecular focuser

Electrostatic hexapoles have been widely used in beam experiments to perform state-selected focusing[45 46 47 48 49]and spatial orientation[50 51 52 53 54] of weak-field seeking, Stark-sensitive molecules. In contrast to those experiments, the hexapole utilized here is quite short in length and functions to increase the OH beam flux by matching molecules from the source into the acceptance aperture of the Stark decelerator. The hexapole is formed by six, hardened steel, 3.175 mm diameter, 50 mm long cylindrical shaped rods, set every 60° at a center-to-center radius of 4.39 mm, mounted to an insulated macor support disk. The rods are mechanically polished and the ends are rounded to a smooth curvature. Alternate rods are electrically connected, thus forming two sets of three rods. Each set is charged to equal magnitude but opposite polarity high voltage. The hexapolar field distribution in the transverse plane allows a weak-field-seeking molecule to be confined and even focused in this plane, leading to transverse phase space "mode-matching" between the supersonic nozzle and the Stark decelerator.

The electric field $|\vec{E}|$ of an ideal hexapole is given as[55]:

$$|\vec{E}| = \frac{3V_o r^2}{r_o^3} \qquad (1)$$

where $V_o$ is the absolute value of the symmetric, opposite polarity voltages applied to each set of rods, $r_o$ is the radius of the hexapole (from the center of the hexapole to the inner edge of the rods), and $r$ is the radial spatial coordinate. A weak-field seeking molecule will thus experience Stark potential energy as: $W_{\varepsilon} = |\mu_{eff} E|$. We define an "effective" dipole moment of the molecule as $\mu_{eff} = \mu \langle \cos\theta \rangle$, where $\mu$ is the magnitude of the electric dipole moment, $\theta$ the angle between the moment and the electric field direction, and ⟨cos θ⟩ represents quantum mechanical averaging over all angles. The radial force $\vec{F}$ is written as:

$$\vec{F} = -\left( \frac{6V_o \, r \, \mu_{eff}}{r_o^{\,3}} \right) \hat{r} \quad . \qquad (2)$$

This linear restoring force results in radial harmonic motion of the weak-field seeking molecules inside the hexapole field region. Thus, in analogy to ray tracing in optics, it is possible to define the focal length $f$ of the hexapole in the thin-lens limit (for $l \rightarrow 0$ as $\sqrt{\frac{6V_o \mu_{eff}}{m r_o^3}} \frac{l}{v}$ remains constant) as:

$$f = \left( \frac{r_o^3}{6\mu_{eff}} \right) \left( \frac{mv^2}{V_o l} \right) , \qquad (3)$$

where $l$ is the longitudinal length of the hexapole, $v$ is the molecule's longitudinal velocity, and $m$ is the molecular mass. Equation 3 demonstrates the focusing strength of a given hexapole linearly increases (decreases) with applied voltage (molecule kinetic energy). For a "real world" hexapole, the finite size of the rods can lead to deviation from Equation (1)[56]. Accounting for this effect, as well as the full Stark shift as illustrated in Fig. 2 (a), detailed numerical simulations of the hexapole focusing effect are shown in Fig. 6. The trajectory simulations deterministically map an initial volume in phase-space to a final volume, which is then matched to the corresponding experimental data by proper weighting of initial molecular numbers. The figure shows three sets of focusing curves consisting of OH signals measured directly after the hexapole, under conditions where the applied hexapole voltage has been pulsed to match to the corresponding input molecules' velocity. Operating the hexapole in such a switched manner is designed to give maximum benefit to a particular velocity class, wherein the hexapole voltages are controlled by fast switches that rapidly charge the rods when the selected molecules enter -- and then terminate the voltages to ground when the molecules exit -- the hexapole region. Thus, molecules with speeds significantly different than the targeted velocity class experience less focusing power of the hexapole, minimizing the aberration effect from the distribution of velocities in the molecular pulse. Symbols in Figure 6 correspond to data points measured for 350 m/s (squares), 385 m/s (diamonds) and 415 m/s (triangles) velocity classes, while the solid lines joining the data represent simulation predictions. From comparison of the simulations to the hexapole focusing data, the transverse temperature of the OH beam is determined to be ~4 K, consistent with a supersonically cooled molecular beam. The upper inset in the figure depicts the contributions from the two weak-field seeking states to the 385 m/s trace (see Figure 2(b)), where molecules in the $^2\Pi_{3/2}$ F = 2, $|m_F| = 0$ and $^2\Pi_{3/2}$ F = 1, $|m_F| = 0,1$ states are marginally focused (dashed line) by the hexapole fields, in contrast to the strong effect experienced by molecules populating the $^2\Pi_{3/2}$ F = 2, $|m_F| = 2,1$ states (dotted line). The two weak-field seeking states are included in the simulations with equal weighting.

This figure demonstrates the powerful molecular-focusing capability of the electric hexapole, as the OH molecules are observed only a few millimeters past the end of the hexapole rods. However, for the deceleration experiment, the hexapole is operated only to provide efficient molecular coupling into the physical opening of the slower. This task requires simply matching the transverse velocity and spatial spreads with the decelerator cross-sectional acceptance and the subsequent requisite focal length is thus longer than that needed to bring the molecules to a sharp spatial focus. Experimentally, we find utilizing ~ ± 7-9 kV applied voltages results in sufficient transverse coupling. This empirical result agrees well with computer simulations of the process. Unless otherwise specified, all subsequent data traces shown in this paper are taken with operating the hexapole in this described pulsed manner.

## Theory and simulations of the Stark decelerator

An initial, well characterized, and relatively high phase-space density source of molecules in the ground state is critical for the most efficient use of Stark deceleration. Supersonic expansion provides a direct and convenient approach towards this goal. The conservative reduction of the pulse's mean velocity is accomplished by exploiting the Stark effect associated with weak-field-seeking molecules. As such a polar molecule moves from a low electric field into a high electric field region, its internal energy increases at the expense of its kinetic energy; thus it *slows down* as it climbs the potential energy hill created by the increasing electric field. If the electric field is rapidly extinguished, the molecule will lose this Stark potential energy. This procedure may be repeated as many times as necessary to lower molecule's speed to any arbitrary final velocity. Critical parameters associated with the deceleration process, such as the number of molecules and the associated velocity spread at the end of deceleration can be determined from a detailed understanding of the deceleration process and the related experimental conditions.

To understand the decelerator, it is necessary to examine the electric field distribution produced by the spatially alternating-oriented electrode stages (see Fig. 3). First, considering the plane perpendicular to the molecular propagation direction, there is a local minimum of the transverse electric field in between a pair of slowing electrodes. Since neighboring stages alternate in their respective orientations, the transverse minimum similarly alternates. Thus, the electrode geometry results in a stage-to-stage, transverse guiding of weak-field seeking molecules, simultaneously as their longitudinal velocity is modified by interaction with the inhomogeneous electric fields. Second, in order to minimize the required number of expensive, fast high voltage switches, all the horizontal (vertical) electrodes of the same polarity are connected to a single switch, thus allowing four switches in total to operate the entire 69 stages of the slower. A consequence of this practical switch minimization is that when the high voltage at a particular slowing stage is removed and simultaneously turned on at the following (and preceding) stage, the local electric field does not go to zero but rather to a value generated by the now active set of electrodes. While this effect reduces the amount of energy per stage that can be extracted by the decelerator, it beneficially maintains the quantum mechanical state identity of the molecules, as the hydroxyl radicals are phase-stably slowed.

Figure 7(a) shows the longitudinal potential energy distribution experienced by OH molecules with our present experimental electrode geometry, generated under normal operating conditions of ± 12.5 kV. The dark line corresponds to the active set of electrodes and the dashed line indicates the other (off) set, where the horizontal axis (spatial phase angle) is defined below. Figure 7(b) demonstrates how switching the electric field between the successive stages leads to a change in the molecular kinetic energy per slowing stage that depends critically on the position of the molecules at the switching time. This dependence is conveniently characterized by utilizing a spatial coordinate $\varphi$, called the phase angle, which is defined at the instant of electric field switching and written as:

$$\varphi = \frac{z}{L} 180^{o} , \qquad (4)$$

where z is the longitudinal position of the molecule and L is the stage-to-stage separation. We define z = 0 ($\varphi$ = 0°) as the point midway between two adjacent stages. To analyze the net effect of deceleration on a spatially distributed pulse of molecules, it is useful to define the *synchronous molecule* as one that is at *exactly* the same phase angle $\varphi_o$ at the moment of each switching of the electric fields; thus, losing an identical amount of energy per stage. While the region 0° < $\varphi_0$ < 180° *does* lead to deceleration of the synchronous molecule, only operating in the range 0° < $\varphi_0$ < 90° results in phase *stable* deceleration of a molecular packet. This can be seen in Figure 7(b), that on the positive slope of the potential hill do molecules which are slightly ahead (i.e. faster) or behind (i.e. slower) the synchronous molecules experience a net restoring force towards the synchronous molecule position. Detrimentally, operating on the negative slope of the hill (i.e., in the range 90° ≤ $\varphi_0$ < 180°) results in faster (slower) molecules becoming progressively faster (slower) relative to the synchronous molecule. By similar arguments, phase stable *acceleration* of a molecular packet is only possible on the positive slope of the potential hill in the range 0° > $\varphi_0$ > -90°. The phase stable ranges are indicated by the shaded portions in Figure 7(b). This restoring force experienced by nonsynchronous molecules when the slower is operated in the proper regime is the underlying mechanism for the phase stability of the molecular packet trapped in moving potential wells. For example, given that the spacing between the successive stages is uniform in the decelerator, if the electric field switching time is also uniform, then the position of the synchronous molecule will be $\varphi_o$ = 0, leading to the so-called "bunched" molecular packet traveling at a constant speed down the slower. For $\varphi_o$ > 0, the proper switching period of the electrodes systematically lengthens from at the beginning of the slower to at the end of the slower.

Defining the excursion of a nonsynchronous molecule from the synchronous molecule position as $\Delta\varphi = \varphi - \varphi_o$ and with the help of a sine function fit to the change of kinetic energy per stage, it is straightforward to write for the time evolution of $\Delta\phi$:

$$\frac{d^2\Delta\varphi}{dt^2} + \frac{W_{\max}\pi}{mL^2}\left( \ Sin(\Delta\varphi + \varphi_o) - Sin(\varphi_o) \ \right) = 0 \qquad (5)$$

where $W_{max}$ is the maximum work done by a slowing stage on a molecule and m is the molecular mass. This equation is the same as the one describing an oscillating pendulum, with an offset equilibrium position ($\varphi_0$). The longitudinal phase-space distribution of the nonsynchronous molecules will thus rotate inside an asymmetric oscillator potential. From numerical integrations of Equation (5), solutions for the stable regions of phase-space are shown as a function of the synchronous molecule phase angle $\varphi_0$ in Figure 8(a). The most relevant feature of this figure is the rapidly decreasing area of stable evolution, defined as the area inside the separatrix. This decrease in phase stable area is easily understood by analogy to a pendulum driven by a constant torque. As the torque increases, the amplitude of stable oscillation decreases. This effect results in an asymmetric oscillator potential for nonsynchronous molecules as shown in Fig. 8(b). As the original phase-space distribution of the molecular source before the decelerator populates the phase stable area for $\varphi_0$ = 0, then as $\varphi_0$ increases, the stable phase-space area decreases, which translates into a decrease in the velocity spread of the molecular packet as well as the number of slowed molecules, setting a practical, observable limit to the ultimate temperature of the molecular packet.

From Figure 7(b), it is seen that for molecules ahead of the synchronous molecule the maximum stable excursion, i.e., the separatrix upper spatial bound, is given as:

$$\Delta\varphi_{\max,+} = 180^o - 2\varphi_o \ , \qquad (6)$$

since molecules with $\varphi_o < \varphi < 180^o - 2\varphi_o$ will be slowed more than the synchronous molecule. A molecule with zero velocity at this position (i.e., at $\varphi = \varphi_o + \Delta\phi_{max}$) will have the maximum stable velocity at the $\varphi_o$ position, thus the longitudinal acceptance of the slower in velocity as a function of $\varphi_o$ is given as:

$$\Delta v_{\max}(\varphi_o) = 2\sqrt{\frac{W_{\max}}{m\pi}(Cos(\varphi_o) - (\frac{\pi}{2} - \varphi_o)Sin(\varphi_o)} \,. \quad (7)$$

Figure 9 usefully summarizes the results of the calculation given above, giving an intuitive expectation of performance under different modes of slower operation. The solid square symbols represent stable areas, A, in phase space, plotted as a function of slowing phase angle $\varphi_o$ and normalized to that of the bunching case ($A_0$). The open circles are plotted along the right axis and depict the maximum longitudinal velocity spread of the phase stable packet. As seen in the figure, operation at increasing phase angles results in fewer but also cooler molecules in the phase stable packet. Hence, with an eye toward experimental goals, a careful balance between the required number of molecules in the phase stable packet and the desired longitudinal temperature must be considered when using the Stark decelerator. While Equation (7) allows for the estimation of the longitudinal temperature of the stable molecular packet, the dependence of the transported molecule number on $\varphi_0$ must be determined. For an initial velocity distribution of $\Delta v_{FWHM}$ (full width at half maximum), the number of molecules successfully loaded into the phase stable bucket $N(\varphi_0)$, normalized to that for the bunching case, is shown to be:

$$N(\varphi_o) \approx (1 - \frac{\varphi_o}{\pi/2}) \frac{Erf(\frac{\Delta v_{\max}(\varphi_o)}{\Delta v_{FWHM}\sqrt{\ln(2)}})}{Erf(\frac{2}{\Delta v_{FWHM}}\sqrt{\frac{W_{\max}}{m\pi\ln(2)}})} \,, \qquad (8)$$

where *Erf* is the Error function. In the above derivation, the area inside the separatrix is approximated as square since we are only concerned with variation relative to molecular bunching. Also, we assume that at the input of the decelerator there is no spatial variation of the molecular packet over the dimension of one slowing stage separation, as well as ignore any convolution of the input distribution due to free flight from the creation region. A more limiting assumption made for the sake of a closed form solution is that the lower separatrix bound relative to $\varphi_o$ varies in the same manner as the upper bound (i.e. $\Delta\phi_{max,-} = -\Delta\phi_{max,+}$). Nonetheless, the results of Equation (8) are in good agreement with the measured molecule number as seen in Fig. 15, where a more detailed numerical solution for Equation (8) that accounts for the proper variation of $\Delta\phi_{max,-}$ is also shown, along with the results of a full Monte Carlo simulation of the slowing process.

The Monte Carlo simulations are performed not only for comparison of the expected stable molecule number, but also more importantly for assessment of the detailed TOF spectra observed, as well as to aid in the understanding of the deceleration process. From the geometrical and electrical properties (sizes, distances, and applied voltages) of the decelerator, a three-dimensional electric field map of the slower is constructed, which when coupled with the understanding of the OH energy structure, gives complete information on the forces experienced by radical molecules within the decelerator. The initial input distribution parameters are determined from both experimental dimensions (spatial parameters) and measurements on hexapole-focused, transversely guided molecular pulses (velocity parameters). The trajectories of four-million molecules (with equal weighting of the two weak-field seeking states) sampled from the initial distribution are calculated for the appropriate Stark manipulation process (i.e., bunching, slowing, or accelerating), starting from the skimmer entrance, then traversing the focusing hexapole and decelerator, and finally reaching the appropriate detection region. All simulation curves displayed in this paper are generated with input parameters of an initial velocity of 385 ± 50 m/s, a transverse spread of ± 32 m/s, and a longitudinal spatial extent of 1.25 cm at the input of the skimmer. We point out the importance of including the lower weak-field seeking state ($^2\Pi_{3/2}$ F = 2, $|m_F|$ = 0 and $^2\Pi_{3/2}$ F = 1, $|m_F|$ = 0,1 levels) to properly reproduce the measured spectra. While its contribution to the signal observed after the hexapole is small (because the molecular focuser more dominantly effects radicals in the $^2\Pi_{3/2}$ F = 2, $|m_F|$ = 2,1 states; i.e., see Figure 6), as molecules with large transverse velocities are progressively lost from the phase stable packet during propagation down the slower, the percentage contribution of the lower weak-field seeking state to the observed signals increases since the molecules in this level entering the decelerator have smaller transverse velocity spread.

## Practical construction and operation of the slower

A particularly important practical issue is the stable operation of high voltages (~ 25 kV) between electrodes spaced only a couple of millimeters apart inside the decelerator and the electrostatic trap. Discharge between electrodes will not only disrupt normal operations of the decelerator, but could also inflict permanent damages to the electrodes. A high vacuum system is helpful in terms of enhancing the threshold for unwanted discharges in the apparatus. Additionally, high vacuum is also useful for reducing OH background collision events. The skimmer acts as an effective differential pumping hole that separates the supersonic expansion source chamber from the decelerator chamber. During quiescence, the source (slower) chamber reaches ~ $10^{-8}$ torr (~$10^{-9}$ torr) (1 torr = 133.3 Pa). Under active conditions, the source (slower) chamber is at ~ $10^{-4}$ torr (~$10^{-7}$ torr).

High voltage components are fabricated from 304L stainless steel, or hardened steel, and subsequently mechanically polished and then electro-polished (for 304L) for the best possible surface finish. These high voltage elements are mounted on insulating materials (macor, alumina, or boron nitride) in such a manner to avoid close proximity to ground or opposite high voltage polarities. When opposite polarity voltage elements are attached to the same insulating support, grooves were fabricated into the insulating surface in order to create longer surface path lengths to minimize current leakage between the components. Once under vacuum, the electrodes are subject to a sequence of high voltage conditioning sessions to ensure stable operation. The conditioning consists of two related but distinct methods: DC current and glow-discharge modes[57], both of which essentially create "microdischarge" events that help clean the surfaces and prepare the electrodes for HV operation. Under DC current conditioning, all electrodes are gradually brought up to 1.2 - 2 times their nominal operating level in incremental voltage steps. A limiting resistor (~ 0.1 GΩ) in series with the HV power supplies limits the possible discharging current to small values. The series current can be actively monitored by a floating digital voltmeter and also via the chamber vacuum pressure. It was empirically discovered that the Bayard-Alpert type ionization gauge that actively monitored the chamber pressure was also inducing low level discharging during the conditioning process. UV lamps have also been utilized to purposefully create these beneficial microdischarges during the current conditioning process. When the apparatus fails to DC current condition to satisfaction, we utilize a sparking glow discharge technique. The chamber is backfilled with a few tens of millitorr of helium gas, and a Tesla coil is used to spark a discharge between the relevant electrodes for approximately ten minutes. The system is then vented and this procedure repeated with a backfill of nitrogen gas. Subsequently, the system is DC current conditioned with significant improvements.

Successful operation of the Stark decelerator relies on thorough knowledge of the initial position and velocity distribution of the OH, and temporally accurate control of the inhomogeneous electric fields in order to ensure stable manipulation. We rely on machining precision in the manufacture of slower components (< ±50 μm stated tolerance) to ensure that any physical discrepancies from the specified dimensions are minimized. The applied electric fields are pulsed using commercially available, fast high voltage switches with a specified ~$10^{-8}$ s rise time. External current limiting resistors are used both to protect the switches from excessive current and to slow down the rise time to ~ 1 μs to minimize the generated electromagnetic interference (EMI). The molecules are insensitive to the switching speed reduction, as even on the μs-time scale, the fastest-moving molecules in the system are almost stationary. We note the importance of reduction of the overall EMI from the system when high voltages are being switched actively.

Two HV power supplies (one devoted to each polarity) generate the requisite high voltage for the slowing stages. Each individual power supply output is coupled in parallel with a ~ 2 μF high voltage capacitor on the input side of the fast switch in order to provide the large currents (12 A peak current) required for the rapid charging of the slowing electrodes. A similar arrangement generates the slightly lower high voltage and current used for the pulsed electric hexapole operation and for the electrostatic trap. A digital delay generator (DDG) acts as the master clock for the experiment, controlling the firing of the pulsed laser, as well as triggering a computer-based timing board. The timing board repetitively generates user-defined timing signals to subsequently drive the fast HV switches that control the OH-generating discharge, hexapole, slowing stages, and trap elements. The timing pattern incorporated into the timing board is derived from computer simulations of the deceleration process as discussed in Section V.

## Phase-space manipulations of the OH molecules: experimental results

*In-situ* experimental observations of OH molecules undergoing longitudinal phase-space manipulation inside the properly timed decelerator provide firm confirmation to the theoretical understanding. Time-of flight spectra from three representative and distinct spatial locations -- before 23 stages, before 51 stages, and in the electrostatic quadrupole trap region -- are selected to demonstrate the evolving character of the phase-stable molecular packet. Where present in all data figures that follow, a downward-pointing arrow indicates the location of the phase-stable molecular packet.

Figure 10 depicts the OH TOF spectra when the decelerator is operated under a phase angle $\varphi_0 = 0°$; i.e., the net longitudinal velocity of phase stable molecules is not changed but the molecular packet maintains tight spatial confinement due to bunching. Unlike the transverse guiding effect shown in Fig. 5, spectra generated under properly timed and pulsed electric fields exhibit peak and valley structures, demonstrating the velocity and spatial selectivity of the inhomogeneous electric fields acting on the evolving molecular packet. The experimental data in Fig. 10 (a), (b), and (c) are plotted with symbols (filled circles) while the full 3-D simulation results are displayed as solid lines. The simulation traces are offset vertically for clarity of presentation. Figure 10 (a) shows the spectra before 23 stages of bunching ($\varphi_0 = 0°$). With such few stages of interaction, longitudinally phase unstable molecules are still dominantly present and detected by the LIF technique, generating the side peak features and contributing to the broad pedestal that the central peak sits on. By the time the molecular pulse has propagated in front of the $51^{st}$ stage under the same experimental conditions, the phase stable packet has become the only distinct feature remaining, as shown in Fig. 10 (b). The molecules that correspond to untrapped orbits in phase-space generate the small, broad background on top of which the phase-stable peak sits. Similarly, the bunched, phase-stable packet dominates the spectrum in Fig. 10(c), where the molecules have exited the slower, propagated a short free-flight distance, and passed through the center of the electrostatic trap region where they are detected.

In principle, all molecules within the selected phase stable packet are transported without loss down the decelerator; the difference in peak amplitude between Fig. 10(b) and (c) is a result of molecular loss due to molecules with too high transverse velocities being progressively removed from the molecular pulse, inefficient coupling between the decelerator and trap region, and scattering from background gas. Figure 10(d) depicts the bunched packet peak areas at different spatial locations from both experimental data and simulations, normalized to the peak before the 23rd stage. This figure, along with the understanding gained from simulations, demonstrates that the apparent loss of signal is primarily due to the 'boiling' away of longitudinal phase unstable molecules as they propagate down the slower, as well as molecules possessing transverse velocities $\geq$ 5 m/s slowly escapes from the phase stable packet throughout the deceleration process. Essentially, the focusing power of the hexapole is boosting the initial total number of molecules coupled into the slower. However, many of these molecules are ultimately not useful as they do not enhance the population of the phase stable packet and are subsequently lost. The simulation results provide excellent agreement with the observed spectra, confirming good understanding and control of the decelerator.

We gain more insight into the slowing process by examining in detail the TOF spectra obtained at a single specific spatial location as the phase angle is varied. Figure 11 focuses on LIF measurements made before the 23rd stage. Figure 11(a) shows the expected broad, basically featureless peak as the packet spreads longitudinally while being transversely guided down the central axis of decelerator under steady-state high voltages applied to all slowing stages. Figure 11(b) depicts the spectra under bunching operation ($\varphi_0$ = 0°), creating the multi-peaked structure, as also shown in Fig. 10(a), wherein the spatial periodicity of the high voltage stages is imprinted onto the OH pulse. The phase angle $\varphi_0$ is then increased to 20°, 40°, 60°, 67°, and 80° in Fig. 11(c), (d), (e), (f), and (g), respectively.

Careful examination of the data traces reveals small changes in amplitudes and subtle time shifts in the location of the peaks as $\varphi_0$ increases. However, the basic structure of the spectrum at this spatial location is mostly unchanged, as the signals from phase unstable molecules dominate the spectra. The downward-pointing arrow indicates the location of the phase stable packet – however, its presence is substantially masked by the signal from unstable molecules. This effect is expected and clearly seen in Fig. 12, where the results of a 3-D Monte Carlo simulation after 22 stages of deceleration (i.e., in front of 23rd stage) are shown for bunching and slowing with $\varphi_0 = 60^o$. Figure 12 represents a snapshot of the longitudinal phase-space distribution of the molecular packet, with each dot corresponding to the longitudinal position and velocity of a simulated molecule. Along the horizontal (vertical) axis is shown the spatial (velocity) projection of the distribution, and a horizontal (vertical) dashed line represents the expected position (velocity) of the phase stable pulse at this instant. Figure12 (a) shows the result of bunching to the 23rd stage, revealing that phase unstable molecules are 'boiling' away symmetrically around the stable molecule bunch. On the other hand, when $\varphi_0 = 60^o$, Fig. 12(b) demonstrates that unstable molecules are lost in only one direction as the phase stable packet is pulled to lower velocities (consistent with the non-synchronous molecule potentials shown in Fig. 8(b)). These phase unstable molecules lead to dramatic effects in the observed TOF spectra of Fig. 11. In the bunching case (Fig. 11(b)), because the unstable molecules are lost symmetrically the bunched peak is only broadened. However, in the case of slowing or acceleration (Fig. 11(c) through (h)), the molecular loss is directional and the center of the observed peak is shifted from the expected location in addition to broadening. From the projection onto the horizontal axis it is evident that a peak does indeed occur at the spatial location expected for the phase stable molecule packet. However, because of convolution with the detection window this peak is unresolved in the large central peak in the TOF data. Also in Fig. 12(b) the manipulation of the input velocity distribution ($V_m$ = 385m/s, $\Delta v_{FWHM}$ ~ 75 m/s) is beginning to become evident as molecules are moved from the central region (385 m/s) towards smaller velocities. The stable molecules *have* been incrementally slowed from the initial 385 m/s (Fig. 11(b)) to 309 m/s (Fig. 11(g)), although this corresponds to only a ~ 40 μs shift in the peak location at this position. We point out that the spectrum is simpler than previously observed (i.e., Fig. 2 in Reference [18]) at this spatial location, when only 14 stages of slowing were utilized, coupled with an additional 9 stages of transverse guidance. Generally speaking, the more stages used, the more straightforward the interpretation of the spectrum will be, as contributions from the interesting though complicated dynamics of the longitudinally phase unstable molecules will be reduced. For display purposes, Fig. 11(g) shows similar characteristics as we select $\varphi_0$ = 80°. At this aggressive slowing phase angle, the selected molecules will actually come to rest before the end of the 69-stage Stark decelerator. Finally, demonstrating the flexibility of this experimental technique, phase stable *acceleration* is depicted in Fig. 11(h), where a negative phase angle is utilized ($\varphi_0$ = -45°). The OH molecules have been accelerated from 385 m/s to 424 m/s. However, similar to the slowing data traces, the acceleration spectrum is complicated by signals from molecules with open trajectories (not contained inside the separatrix) in longitudinal phase-space.

We next observe molecular packets after 50 stages of operation. Similar to above, Fig. 13(a) shows the TOF spectrum solely under continuous, transverse guidance fields. From Fig. 14 it is clear that after 50 stages of operation (i.e. observed in front of the 51$^{st}$ stage) essentially all of the phase unstable molecules have boiled out of the phase stable region. This significantly simplifies the interpretation of TOF spectra since only the well-characterized stable molecules remain in the peak of interest. Indeed, under bunching operation experimentally (i.e., $\phi_0$ = 0), Fig. 13(b) shows a single tall central peak representing the phase stable molecules that are maintained together with their net longitudinal velocity unchanged. The much smaller, lower-lying broad feature again arises from the phase unstable molecules. As the slowing phase angle is further increased, the behavior of the decelerated packet becomes manifestly clear in the data at this spatial location unlike after 22 stages. The slowed peak corresponding to the phase-stable packet evolves towards later arrival times, demonstrating its steadily decreasing velocity. In Fig. 13(c), (d), (e), (f) and (g) , the slowed peak appears at later time positions as the velocity of the OH molecules is reduced to 337 m/s, 283 m/s, 220 m/s, 198 m/s, and 168 m/s, respectively. As expected, the peak amplitude decreases as the phase angle gets larger, reflecting the reduced stable area in phase-space. Again, from Fig. 14(b) for $\varphi_0$ = 60°, we see this expected behavior in the snapshot projections of the longitudinal phase space distribution. In the horizontal projection we observe a single phase-stable peak that has been moved towards the rear of the molecule pulse, while in the velocity projection we see that the input velocity has been significantly modified, resulting in a single packet of decelerated molecules traversing the slower.

Interestingly, there are distinctly different spectra features generated under bunching versus slowing conditions at this spatial location. In contrast to in front of the 23$^{rd}$ stage where the spectra maintain essentially the same characteristics at all operating phase angles, after the 50$^{th}$ stage at any non-zero phase angle, the spectrum changes from a single peak to a distinct triplet of peaks, where one of these structures corresponds to the phase stable packet of molecules that shifts to later (earlier) times for stable deceleration (acceleration) in the TOF observation. Within the presented data's time resolution, the other two peaks remain essentially at the same temporal locations for the same sign of phase angle, but these positions undergo a large time shift for opposite sign phase angles, as seen by comparison with Fig. 13(h). These structures may result from phase unstable molecules hopping around neighboring traveling potential wells.

We now examine OH signals observed beyond the final ($69^{th}$) stage of the decelerator and moving within the trap region. The electrostatic trap consists of three elements which produce a quadrupole-configuration confinement field in order to spatially contain sufficiently slowed, Stark-sensitive molecules in three dimensions. A stainless steel ring with 2 pairs of orthogonally-oriented 3 mm holes for fluorescence detection forms the central axis of the trap, located ~14 mm from the middle of the final slowing stage. Two identical end cups with 2 mm apertures along the decelerator beam axis create the symmetric end-caps of the trap. The trap element surfaces -- the inner ring and inner end-cap sides -- are machined to a specific curvature from 304L stainless steel and held in precise orientation and separation by insulating support rods. The three pieces are electrically isolated both from earth ground and each other, enabling voltages to be applied independently for user-defined, precisely timed, molecular loading and trapping sequences. In order to provide high-efficiency molecular fluorescence detection within the trap region, an in-vacuum, 25.4 mm diameter lens is mounted one focal length away (f = 35 mm) from the trap center, axially-aligned to one pair of the ring element apertures. A second identical relay lens is mounted 2f away (70 mm) to transport the fluorescence photons, forming an image of the trap central region at the photocathode of an external PMT for molecular detection. While the back end-cap is always grounded, high voltage of -12.5 (-9.0) kV is applied to the central ring (front end cap) during the loading sequence. Subsequently, the front end-cap is terminated to ground, resulting in an eventual trap volume confinement of ~ 1 $mm^3$.

While we have recently observed the first signatures indicating OH radicals slowed to ~27 ± 6 m/s are successfully loaded into the trap, and spatially confined by the high electric fields, this topic is beyond the scope of the present paper and will be presented in a separate, future publication. We present here further TOF spectra of OH molecules observed moving with varying velocities through the trap region after the decelerator, which is operated under various conditions. Where appropriate, the simulation results are plotted as lines that have been offset vertically from the data trace for viewing clarity. Figure 15(a) shows the transversely guided pulse of OH molecules in the trap region with a mean pulse velocity of 385 m/s. Under bunching conditions selected for this mean velocity, the expected narrow, single, phase-stable packet shows up as a single large peak depicted in Fig. 15(b), arriving at 1.37 ms. In Fig. 15(c) at a phase angle $\varphi_0$ = 45°, the slowed peak arrives later in time as the molecular speed is further slowed to 210 m/s. In fact, the peak has been moved completely off of the residual signals from phase unstable molecules that form the broad pedestal in Fig. 15(b). This complete separation occurs at a smaller value of $\varphi_0$ than that for 51 stages since the entire 69 stages of decelerator are utilized for kinetic energy extraction. Finally, phase-stable acceleration is shown in Fig. 15(d), wherein the molecular packet is accelerated to 541 m/s, arrives correspondingly earlier in the TOF spectrum and creates the narrow peak at 1.19 ms. Interestingly, similar to the observations after the $51^{st}$ stage, at non-zero phase angles, the double peak structure appears to persist in the data even after an additional 18 stages of high electric field manipulation. Above the data traces, simulation spectra with higher temporal resolution are plotted. Consistency between simulation and measurement is quite good, with locations of the phase stable packet and relative peak heights demonstrating excellent agreement.

We now focus with higher temporal resolution on measuring solely the phase stable packets generated by the decelerator moving through the trap region. Successful operation of the decelerator relies on precise knowledge of the molecular packet's velocity and location in order to be able to stably slow the packet. Hence, we can easily predict when the pulse of molecules will arrive in the trap region after it exits the decelerator. We note, that at low phase angles, when the velocity change of the molecules is not that large, the analytic expression (i.e., sine fit) to the potential energy hill presented in Fig. 7(b) enables sufficient accuracy to predict the peak arrival times. However, at higher phase angles, the deviation between the analytical calculation and the actual shape of the potential becomes a significant factor such that a numerical determination of the potential must be used to generate the proper time sequence to operate the decelerator. In fact, both simulations and measurements demonstrate that the molecular packet actually becomes phase unstable once $\varphi_0$ becomes greater than 40°, unless the more accurate calculation of the time sequence is used. Finally, as a minor detail, when the molecules are exiting the decelerator through the last stage, the voltage on the last electrodes is switched off, while the voltage on the preceding stage is turned on. The presence of this non-zero electric field results in a small accelerating force to the molecules. This effect needs to be taken into account to determine accurately how the molecules propagate from the last decelerator stage through the essentially field-free region into the trap.

In Fig. 16, we detect the molecules with 2.5 μs timing resolution per point. Phase stable packets are measured over a broad range of operating parameters, both in acceleration, bunching, and slowing for $-70° \leq \varphi_0 \leq 65°$. Fig. 16 traces (a), (b), (c), and (d) represent OH radicals that have been accelerated from an initial velocity of 385 m/s to a maximum speed of 545 m/s. The phase stable molecules arrive earlier in time relative to the bunched peak (Fig. 16 trace (e)) and create the observed increasingly narrow structures. The bunched packet corresponds to $\sim 10^4$ molecules at a density of $\sim 10^6$ $cm^{-3}$. Traces (f), (g), (h), (i), (j), (k), (l), and (m) show phase stable OH packets which have been decelerated to correspondingly lower velocities, as indicated in the figure. As the molecular packet is increasingly slowed, the molecules arrive progressively later in time. For $\varphi_0 >$ 40°, the slowed peak is completely separated from the broad background signals produced by residual phase unstable molecules. In contrast to the accelerated molecules, the slowed, stable molecular packets broaden as the phase angle increases, due the reduced molecular speed.

As the molecules slow down towards rest, there is an artifact in detection of molecular numbers due to the finite spatial size of the detection window. Basically, slow molecules are counted multiple times as they move through the detection region where the LIF light pulses are shifted in fine time steps to obtain high resolution TOF spectra. When this effect is properly de-convolved from the actual underlying signal size, the number (N) of OH molecules in a phase stable packet as a function of phase angle can be determined. The measured results (filled circles), normalized to the bunching peak number $N_0$ are plotted in Fig. 17. In addition, points generated under Monte Carlo simulations (open diamonds) as well as theory curves from the simple analytic expression of Eq. (8) (solid line) and the more accurate numerical calculation (dotted line) are shown. The agreement between experimental results and simulation is quite good. The main sources of error are peak reproducibility and accurate estimation of the peak area of the bunched packet as it is temporally coincident with signals from phase unstable molecules. Disagreement between the measured data points and the simulation results are within experimental error. Deviation of the data points under acceleration conditions ($\varphi_0 < 0°$) are primarily due to insufficient time resolution in the data points, making peak area estimates difficult.

## Conclusions, future directions, and other discussions

The work reported in this paper has centered on providing cold hydroxyl radicals through the use of the Stark deceleration of a supersonic beam. Specifically, our work from both experimental and modeling perspectives has been focused on uncovering the complex dynamics, which govern the evolution of the molecules within the decelerator, as well as efficiently produce molecules for trapping and subsequent study. In our current experiments we accelerate/decelerate a supersonic beam of free radicals to a mean speed adjustable between 550 m/s to rest, with a translational temperature tunable from tens of milliKelvin to 1 K. These velocity-manipulated packets contain $10^{3-}$ [6] molecules (depending on the translational temperature) at a density of ~$10^{4-6}$ $cm^{-3}$ inside the moving potential wells.

Improvements will come from several different fronts. One of the most important future developments is to enhance the number and phase space density of the initial supersonic beam expansion source. Since a Stark decelerator cannot enhance the phase space density, progress along this line will be particularly important in order to push the low temperature limit of the cold molecular samples. We will explore improvements to the existing system as well as alternative techniques such as photolysis to attempt to produce colder and more intense OH beams.

Of course, one of the most effective, demonstrated approaches to enhance the phase space density is to laser cool the sample, especially when the sample density is not sufficiently high or the ratio of elastic vs. inelastic collision rates is not favorable to afford direct evaporative cooling. For OH radicals, it will be interesting to explore laser cooling effects through specific, relatively radiatively closed electronically excited pathways. For example, due to the large Franck-Condon overlapping factor, OH molecules excited from the v" = 0 level in the ground electronic state to the v' = 0 level in the first electronic state will return to the same ground state at ~ 99.3% branching ratio. The cooling force associated with this electronically allowed transition is comparable to laser cooling of alkali atoms. Within this 0 – 0 transition band, with strict selection rules ($\Delta J = 0, \pm 1$; $\Delta J = 0, \pm 1$; and + ↔ - for parity operation), the slowed and trapped molecules in the (J =3/2, "f", F = 2) state would need only two independent repumping lasers, in addition to the cooling laser, to close the cooling cycle. There are actually five leakage states, however, small frequency gaps between hyperfine levels can be bridged using electro-optical modulators. Although we still have a 0.7% un-repumped leakage to other vibration levels, when the molecules are already prepared at a few tens of millikelvin temperature, noticeable cooling effects should take place.

Because of the existence of an unpaired electron, the OH molecule also possesses a large magnetic moment (~1.4 $\mu_B$) and is suitable for magnetic trapping. We are exploring a hybrid trap configuration combining both electrostatic and magnetic fields for trapping of cold OH. Such a trap will provide an ideal environment to study cold, intermolecular interactions with a varying electric field to influence the molecular dipoles, as well as cold collisions between atomic and molecular samples.

We are grateful to H.L. Bethlem and Gerard Meijer for fruitful interactions and technical advice. We also thank C. Lineberger, S. Leone, and J. Daily for equipment loans, and J. Bohn and D. Nesbitt for useful discussions. We acknowledge important help we have received from A. Pattee and T. van Leeuwen. This research work is supported by the Keck Foundation, NSF, and NIST. H. J. L. acknowledges support from N.R.C.

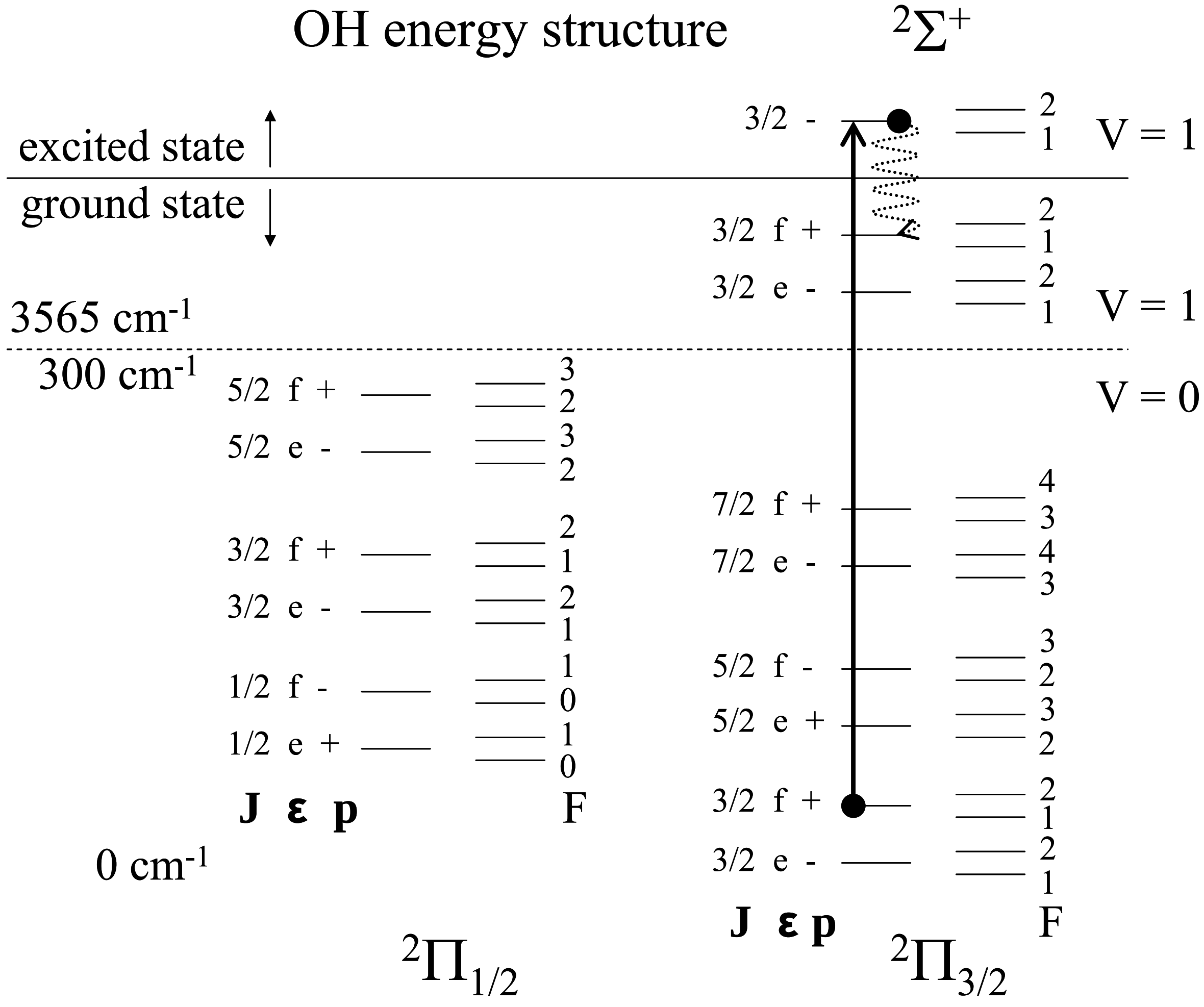

Figure 1: Relevant energy level structure of OH, showing the first electronically excited state and the first two vibrational levels in the electronic ground state. ε indicates symmetry ("e" and "f" states) from λ-doubling and p indicates overall parity. The transition (solid arrow) and decay (wavy, dotted arrow) pathways used for laser induced fluorescence detection of OH are also indicated.

Figure 1 Bochinski et al.

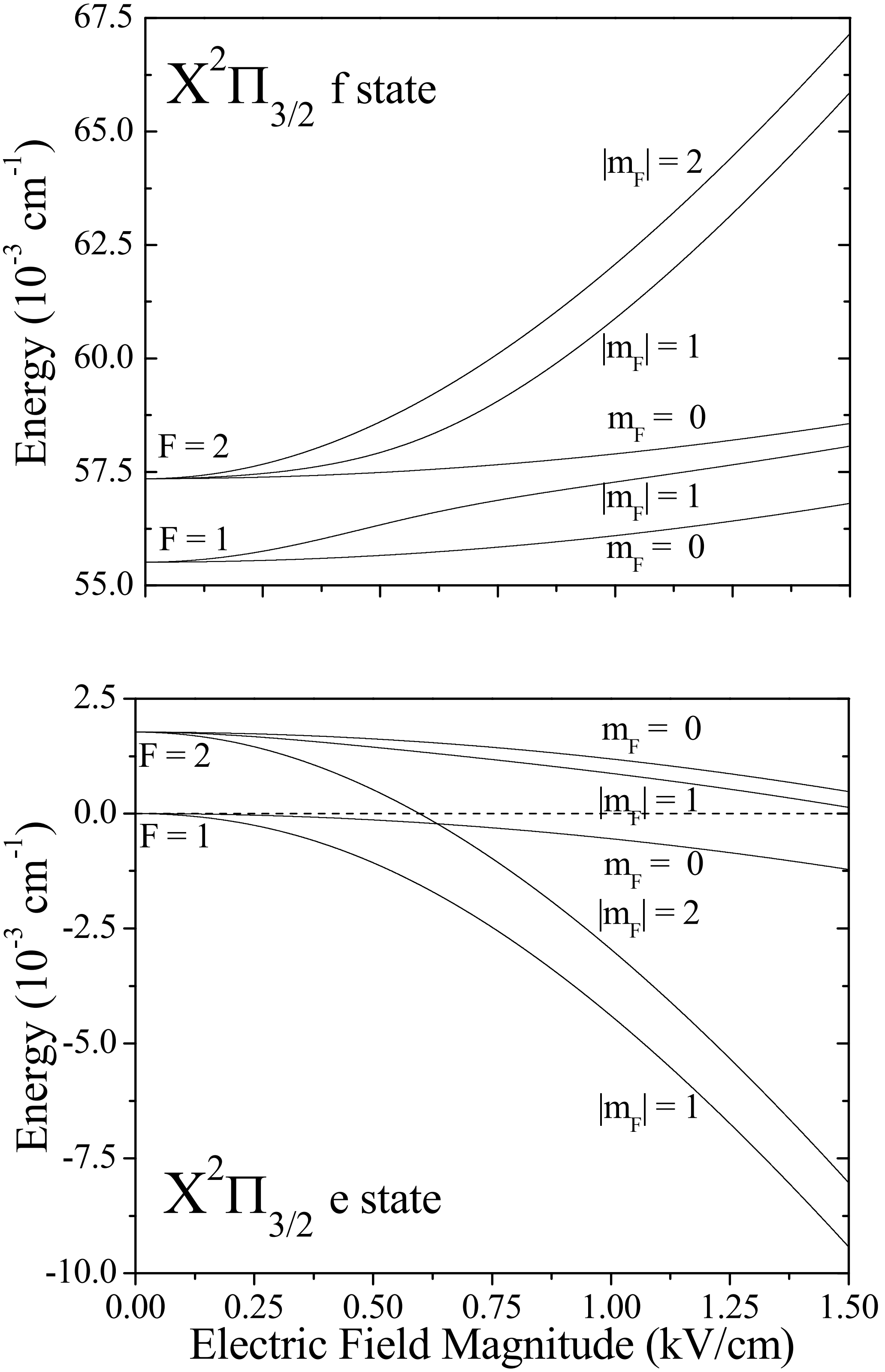

Figure 2 (a) Bochinski et al.

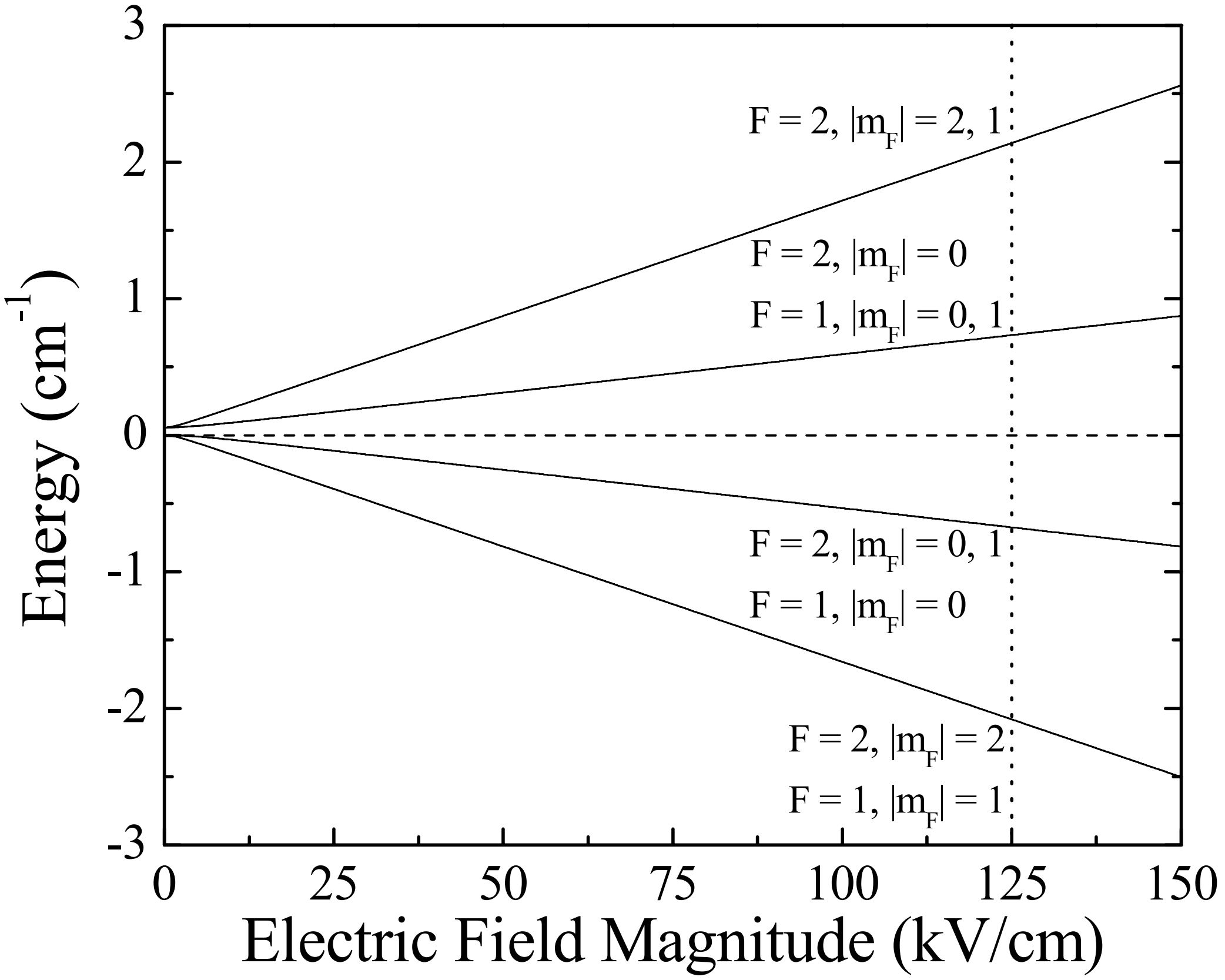

Figure 2: (a) The Stark energy shift for ground state OH molecules in low electric fields. The initial e (f) state is shown in the lower (upper) panel, where the hyperfine levels are indicated next to the traces. (b) The Stark shift for ground state OH molecules in high electric fields. The vertical dotted line indicates the regime where the Stark decelerator operates; the dashed horizontal line denotes zero energy.

Figure 2(b) Bochinski et al.

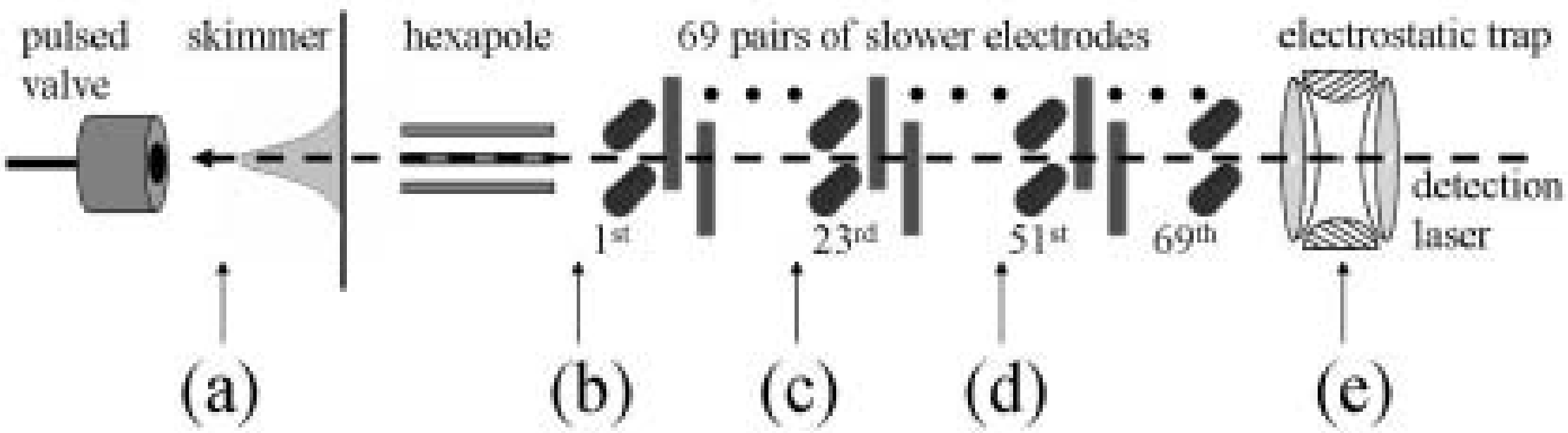

Figure 3: System schematic of Stark decelerator, displaying the pulsed valve, the molecular beam skimmer, the electric hexapole, the electrode stages, and the trap region. The electrode stages alternate orientation (vertical – horizontal) as shown in the figure. The spatial locations indicated by arrows correspond to the measured data traces in Figure 5.

Figure 3 Bochinski et al.

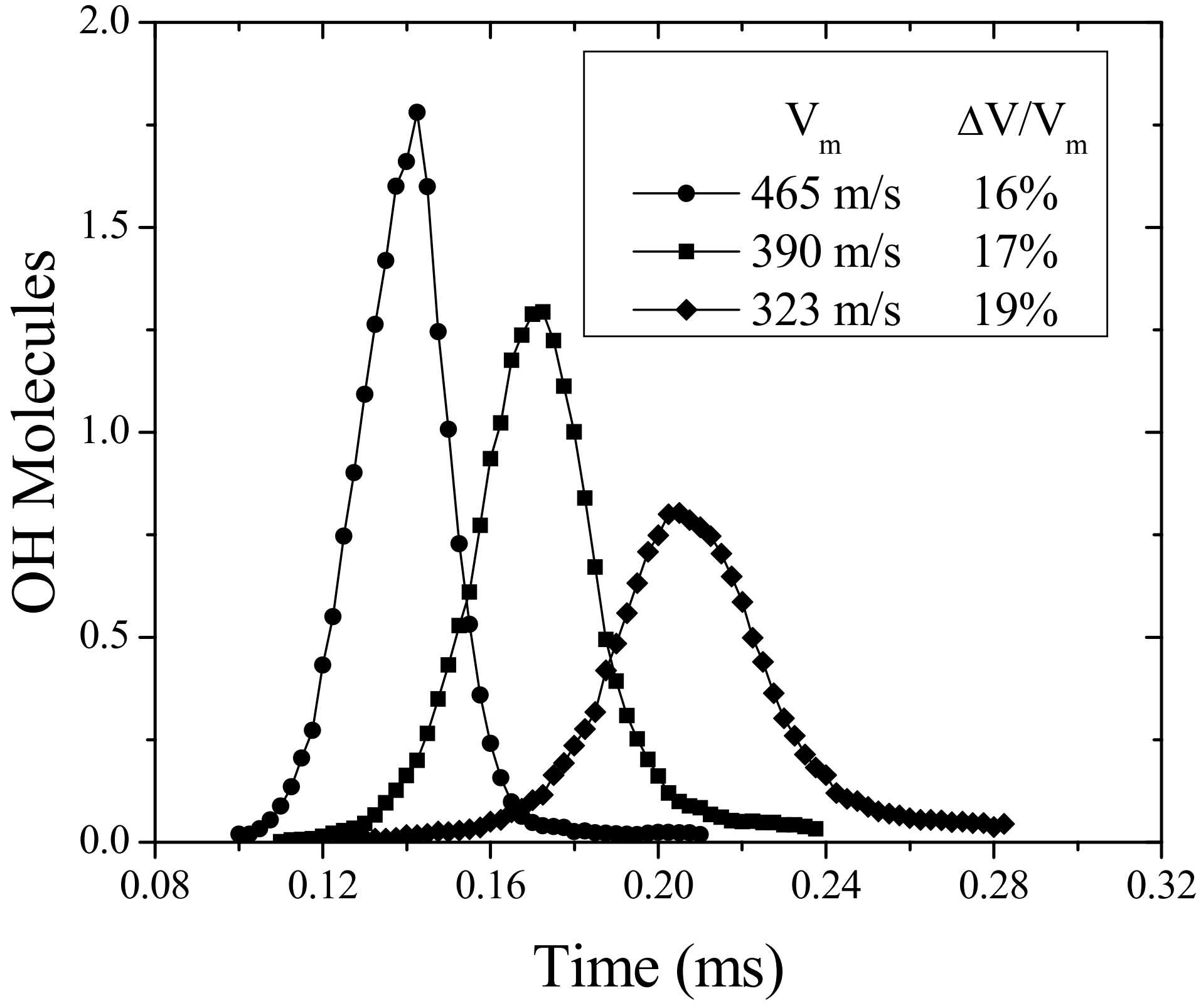

Figure 4: Controlled tuning of the initial mean molecular speed; varying the discharge initiation time creates OH pulses with differing mean speeds ($V_m$), FWHM velocity widths ($\Delta V$), and amplitudes.

Figure 4 Bochinski et al.

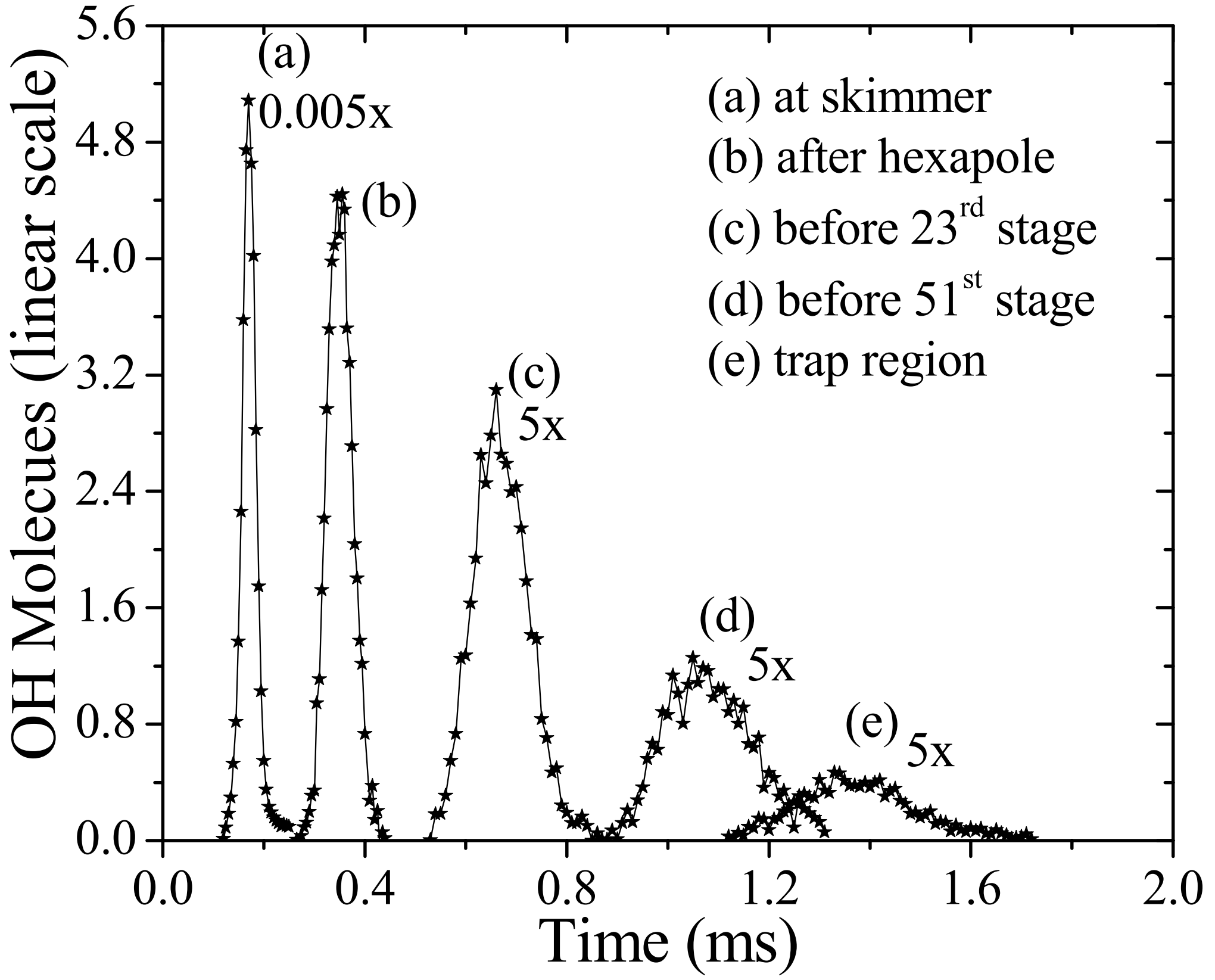

Figure 5: TOF LIF measurements of the propagation of an OH molecular pulse (a) before skimmer, (b) after hexapole, (c) before the 23rd stage, (d) before the 51st stage, and (e) in the trap region. Traces (c), (d), and (e) are taken with the hexapole operated as described in Section IV and constant, transverse guidance voltages applied to the slower.

Figure 5 Bochinski et al.

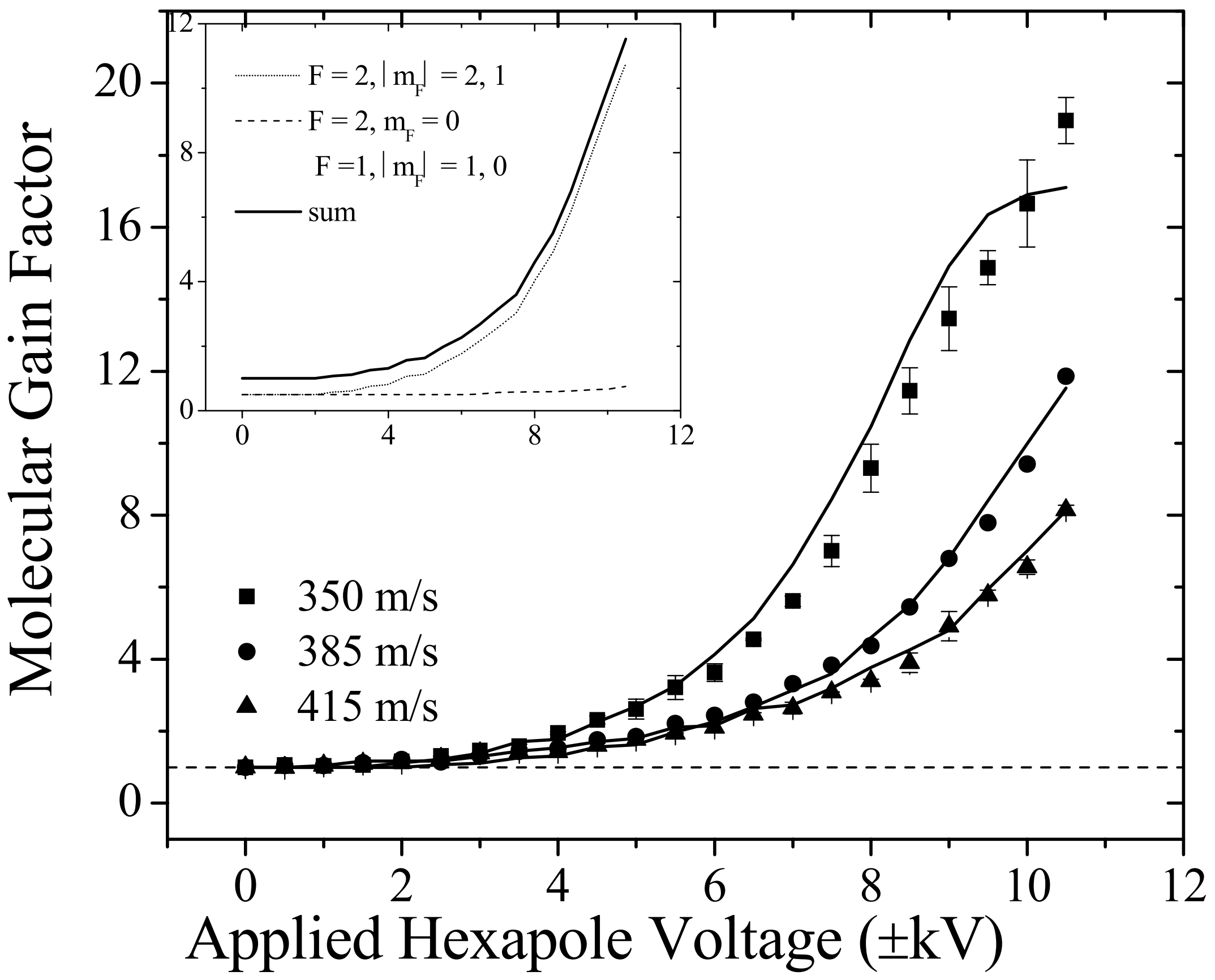

Figure 6: Hexapole focusing curve for different OH molecular velocities. The symbols represent the data points for molecules with velocities of 350 m/s (squares), 385 m/s (diamonds), and 415 m/s (triangles), while the solid lines represent the corresponding simulation result. The inset depicts the contribution of the $^2\Pi_{3/2}$ F = 2, $|m_F| = 2, 1$ states and the $^2\Pi_{3/2}$ F = 2, $|m_F| = 0$ and $^2\Pi_{3/2}$ F = 1, $|m_F| = 0, 1$ states
to the observed signals for the 385 m/s trace.

Figure 6 Bochinski et al.

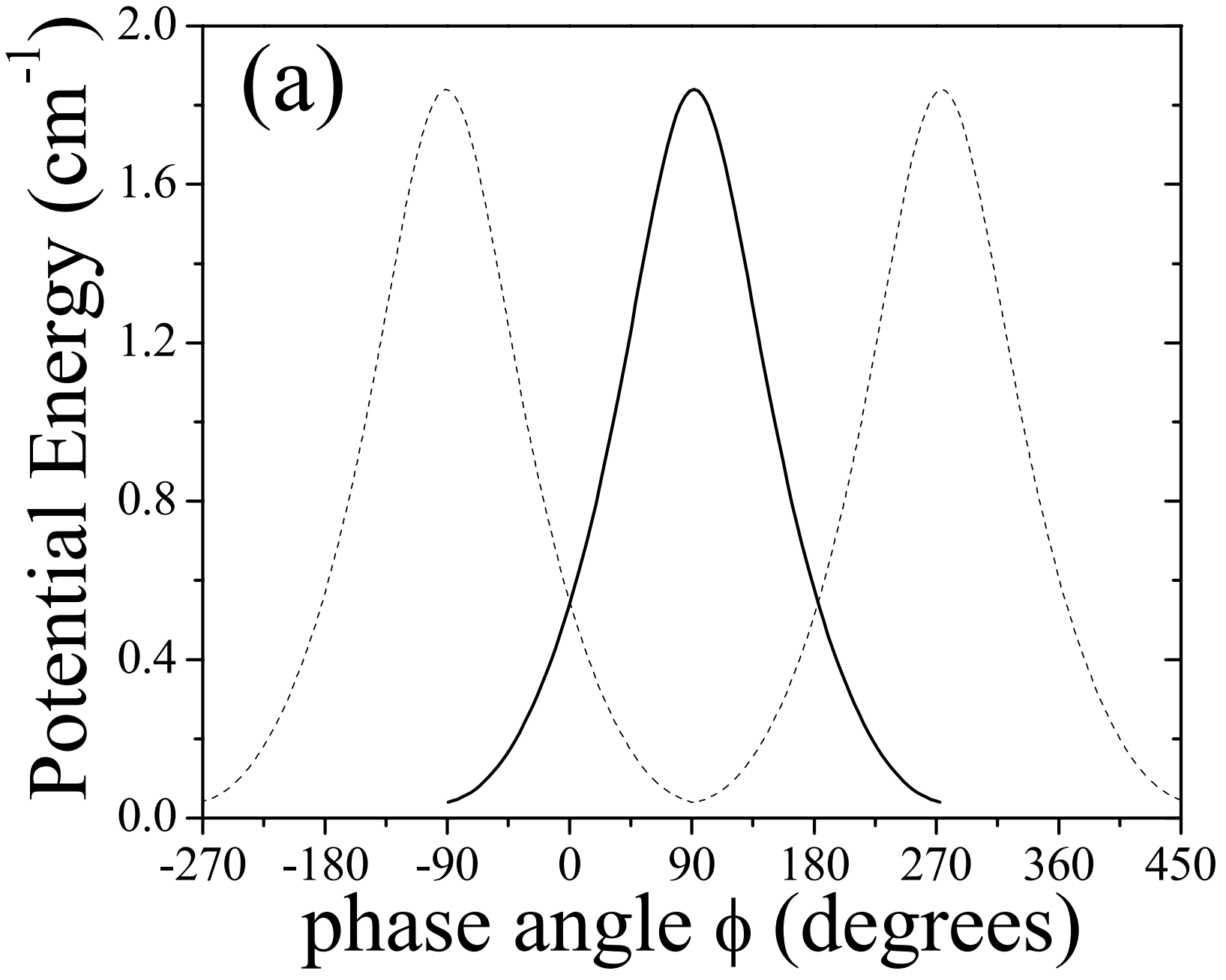

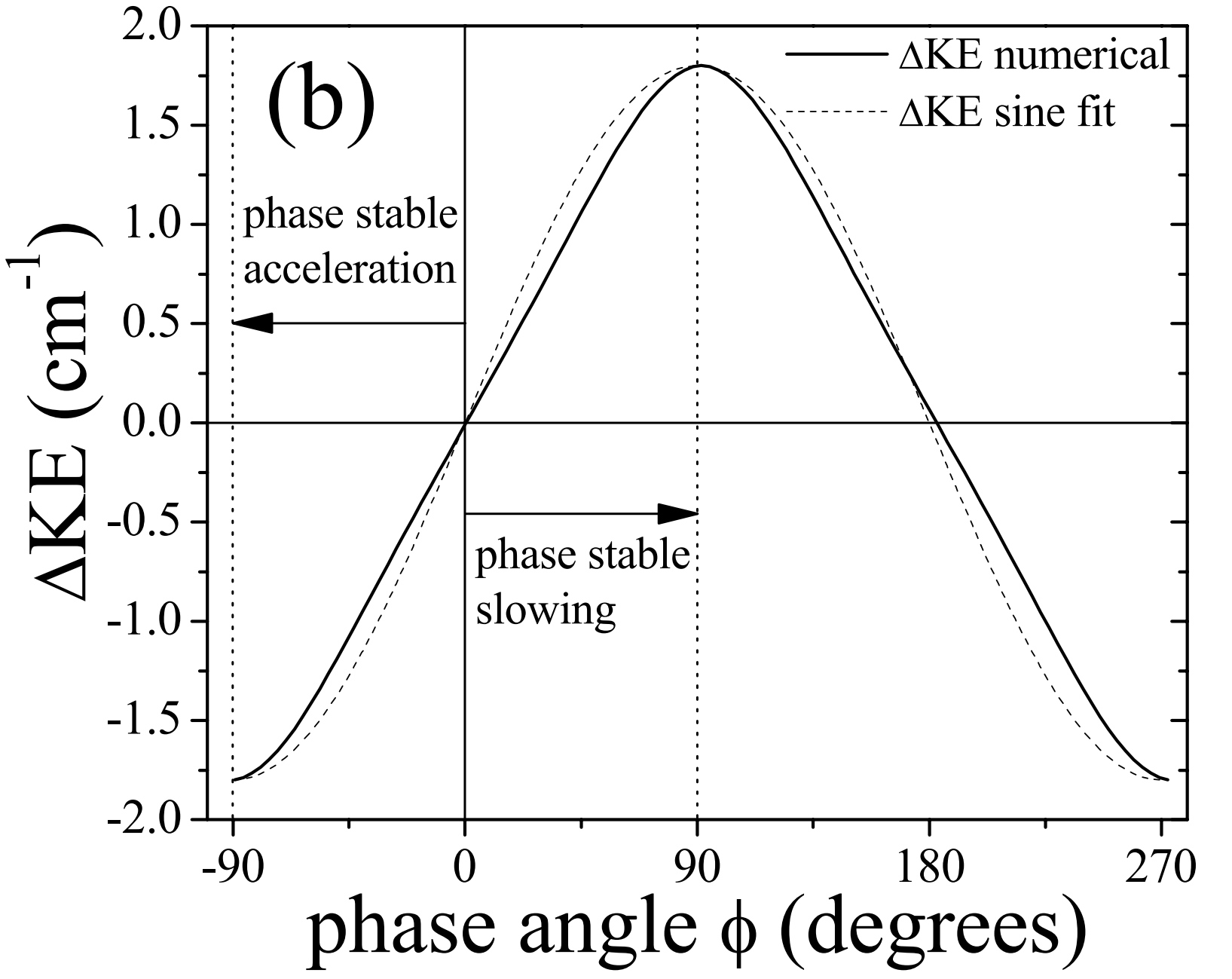

Figure 7: Phase stable operation of the decelerator. (a) Longitudinal Stark energy potentials generated by the two sets of electrodes, where the solid (dotted) line represents the potential from the active (grounded) set of electrodes. (b) Kinetic energy loss per stage (ΔKE) experienced by molecules from switching between the two potentials given above. Solid (dashed) line corresponds to a numerical calculation (sine function fit).

Figure 7 Bochinski et al.

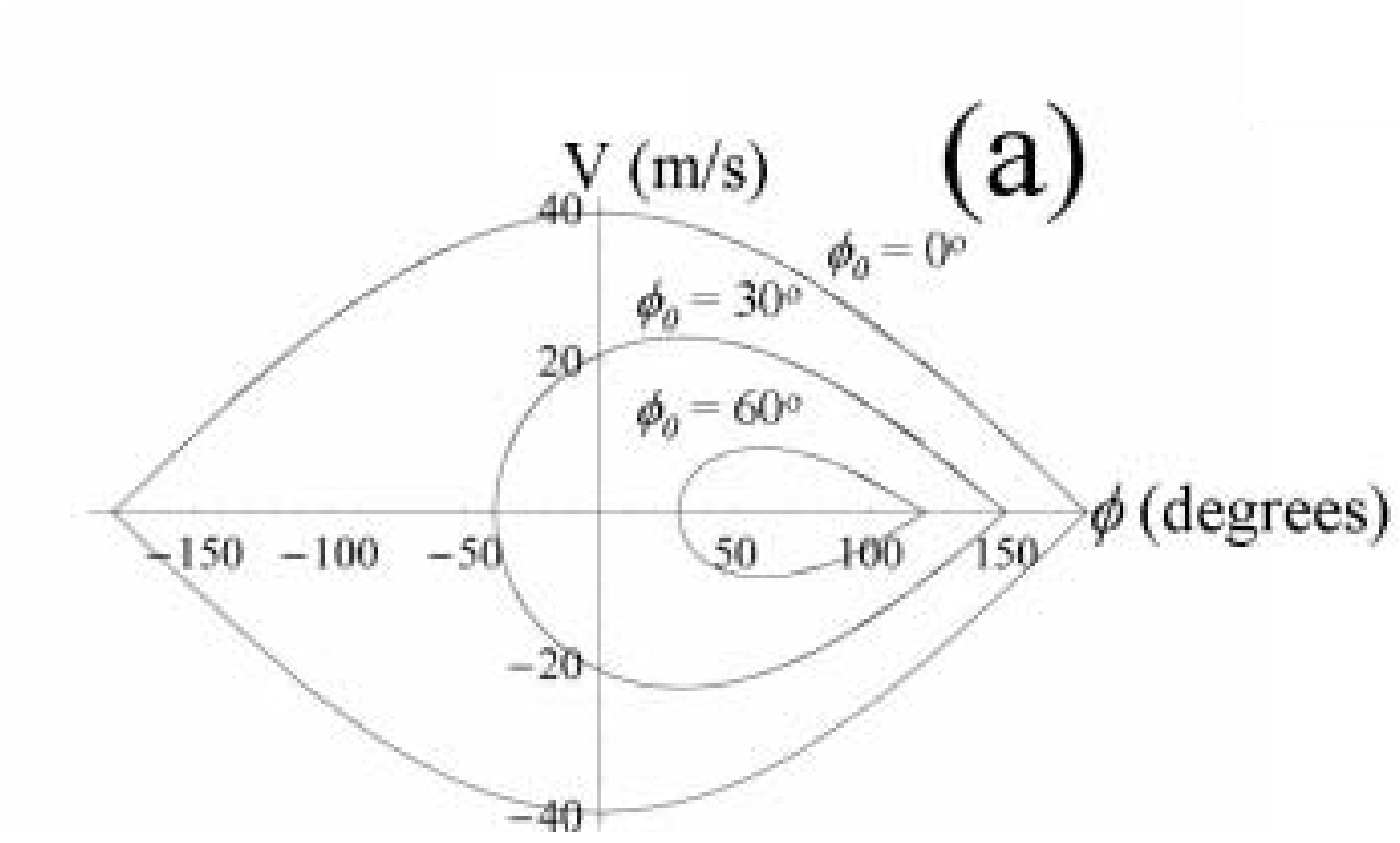

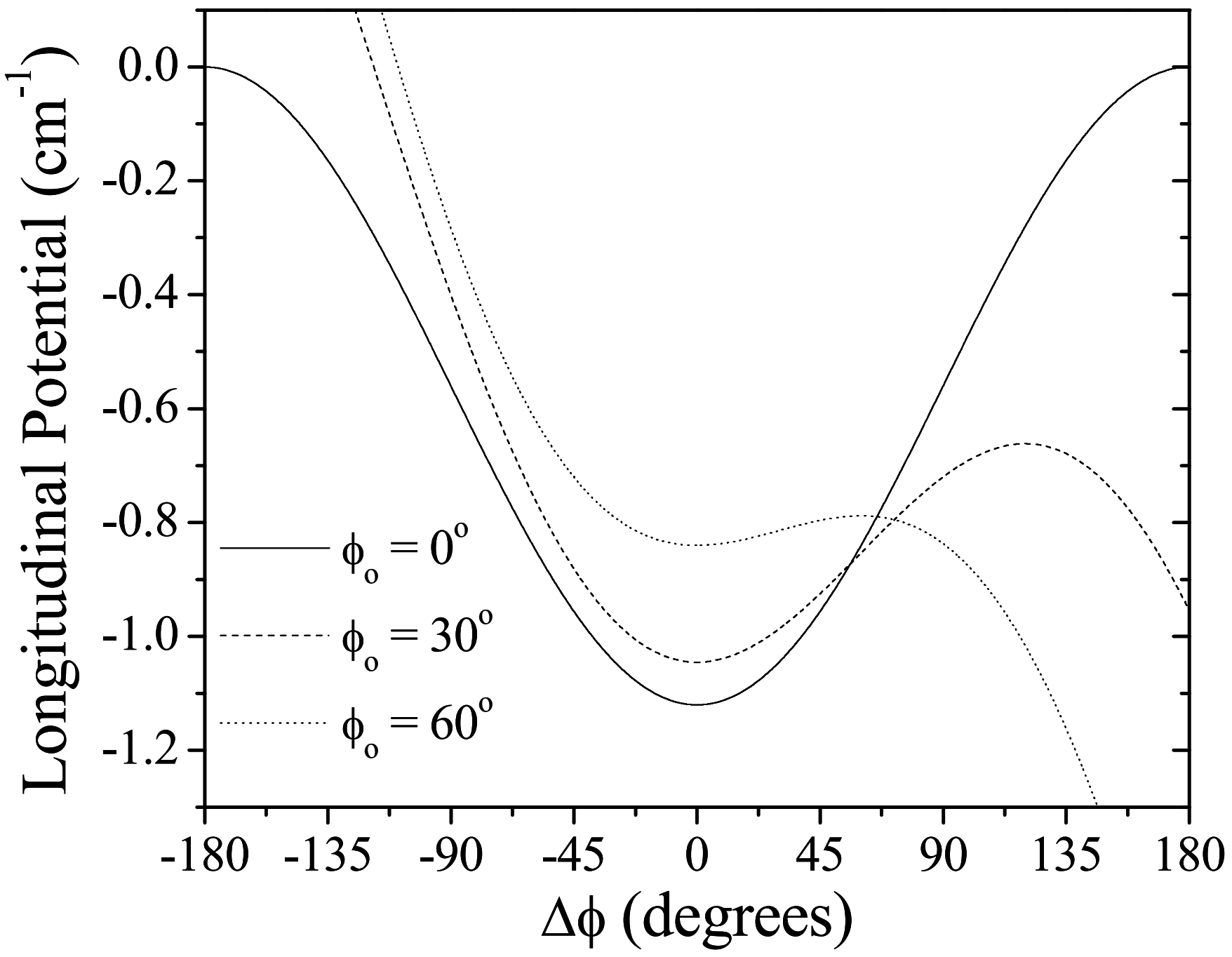

Figure 8: (a) Phase stable area (separatrix area) versus phase angle for $\varphi_0 = 0°$, 30°, and 60°. As the phase angle is increased, the stable area is reduced; equivalently, the stable longitudinal velocity width narrows, decreasing the numbers of molecules in the phase stable packet. (b) Longitudinal traveling potential wells at various phase angles of $\varphi_0 = 0°$, 30°, and 60°.

Figure 8 Bochinski et al.

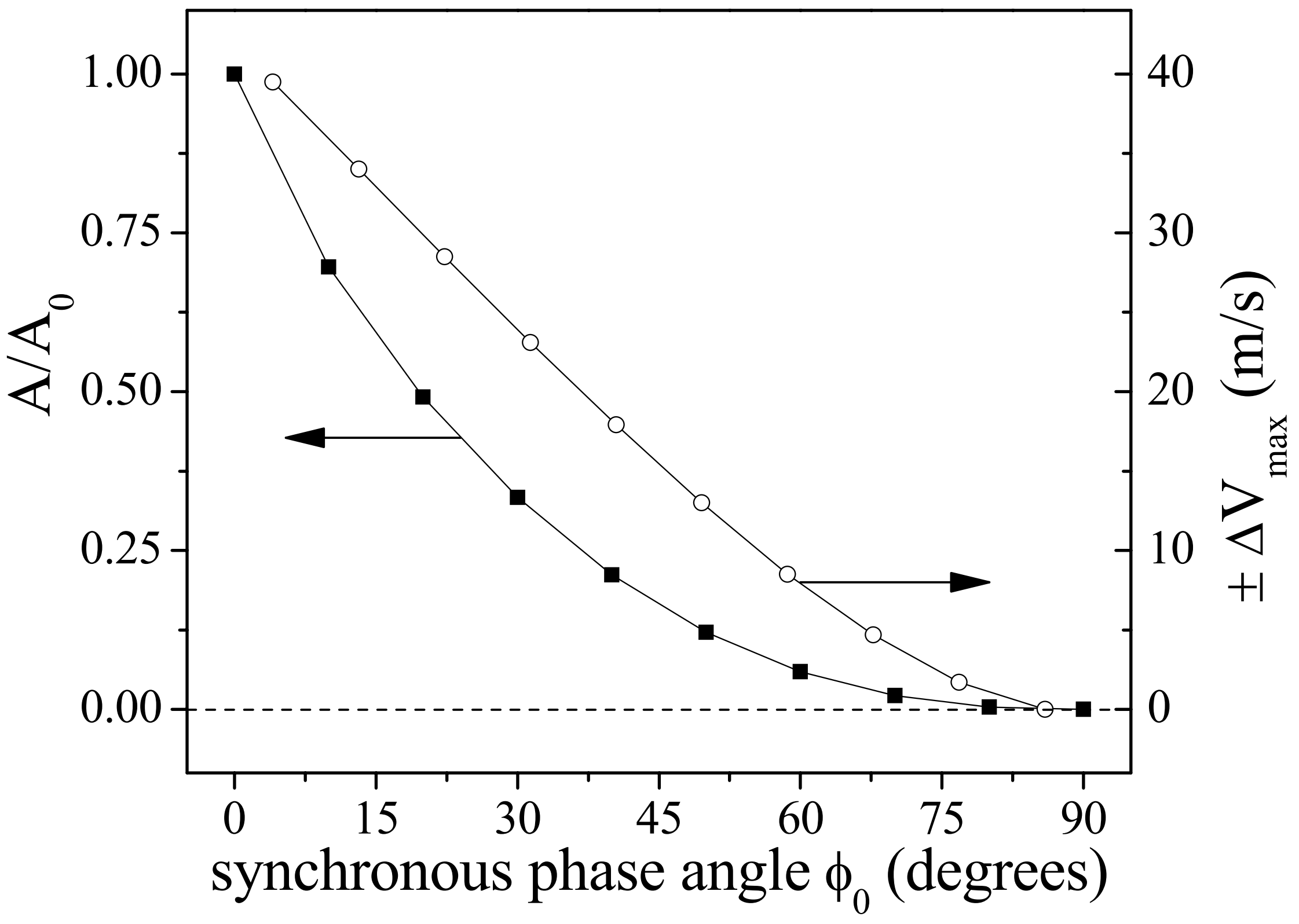

Figure 9: Phase stable area (separatrix area) A versus slowing phase angles $\varphi_0$, normalized to the phase stable area under bunching operation $A_0$. The subsequent maximum allowed velocity spread of the phase stable packet $\Delta V_{max}$ is shown with respect to the right axis.

Figure 9 Bochinski et al.

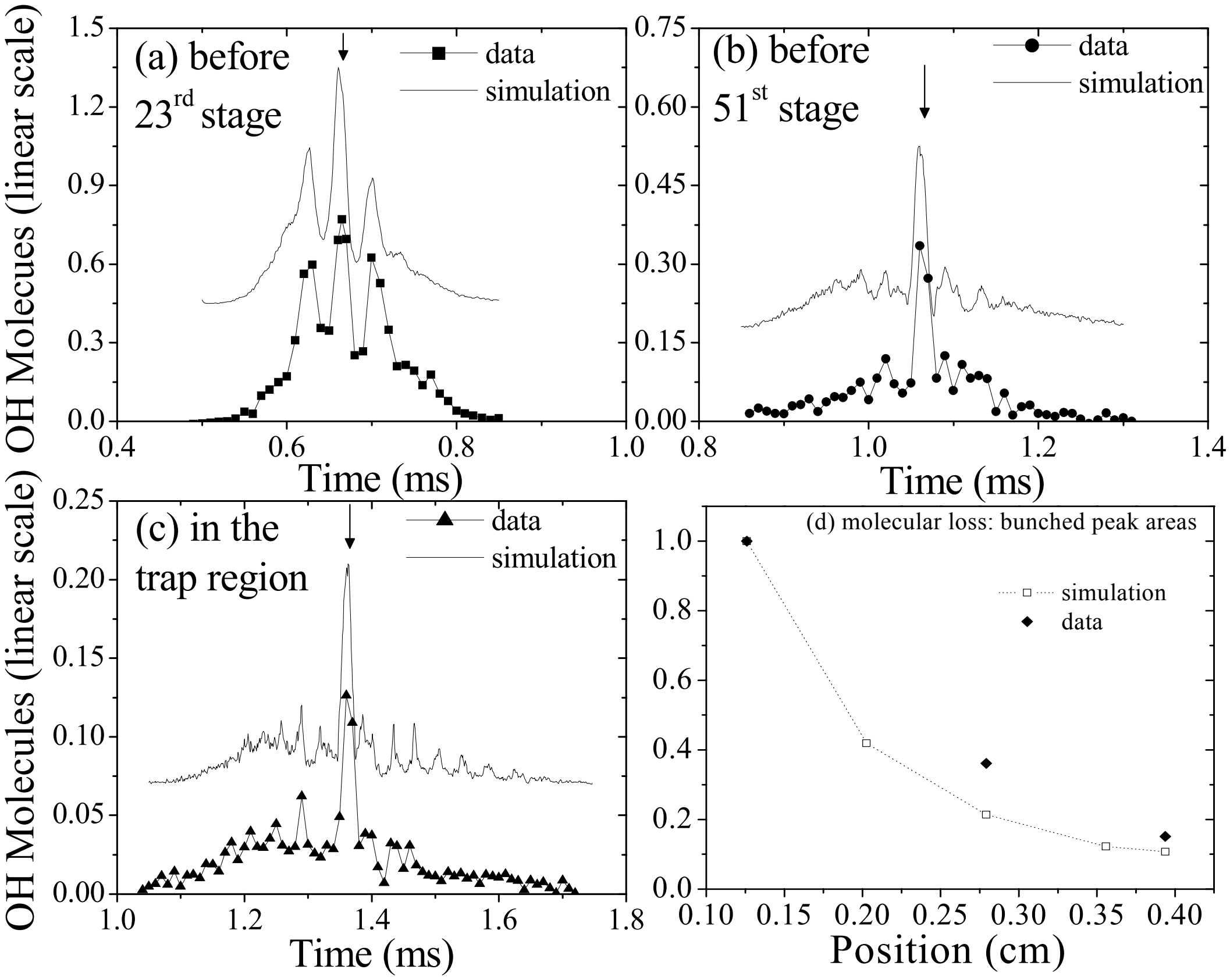

Figure 10: OH LIF TOF spectra under bunching ($\varphi_0 = 0°$) decelerator operation (a) before the 23$^{rd}$ stage, (b) before the 51$^{st}$ stage, and (d) within the electrostatic trap region. The data points are drawn as squares, circles, and triangles while the simulation results are shown as solid lines, with a vertical offset for viewing clarity. Figure 10(d) shows both experimental and simulation results on bunched peak areas, normalized to the 23$^{rd}$ stage; the reduction in peak height with position is due to longitudinally phase unstable molecules as well molecules with radial velocities ≥ 5 m/s being lost from the packet.

Figure 10 Bochinski et al.

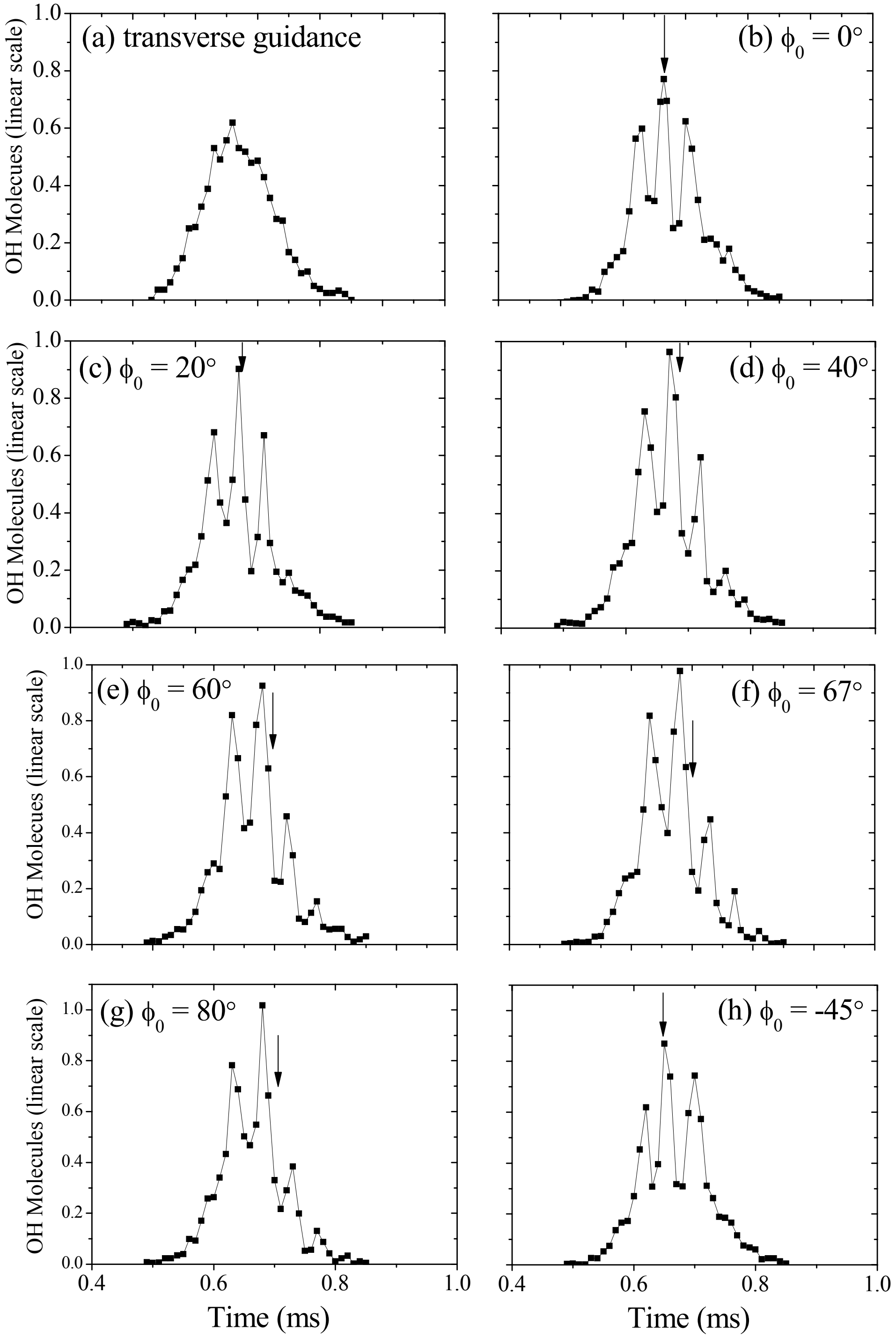

Figure 11: OH packet of molecules in front of the 23rd stage under (a) transverse guidance, (b) $\varphi_0$ = 0° [385 m/s], (c) $\varphi_0$ = 20° [365 m/s], (d) $\varphi_0$ = 40° [344 m/s], (e) $\varphi_0$ = 60° [323 m/s], (f) $\varphi_0$ = 67°[317 m/s], (g) $\varphi_0$ = 80° [309], and (h) $\varphi_0$ = -45° [424], where the number in square brackets indicates the mean velocity of the phase stable packet. At this spatial location, under all operating conditions the spectra are dominated by signals from still present phase unstable molecules.

Figure 11 Bochinski et al.

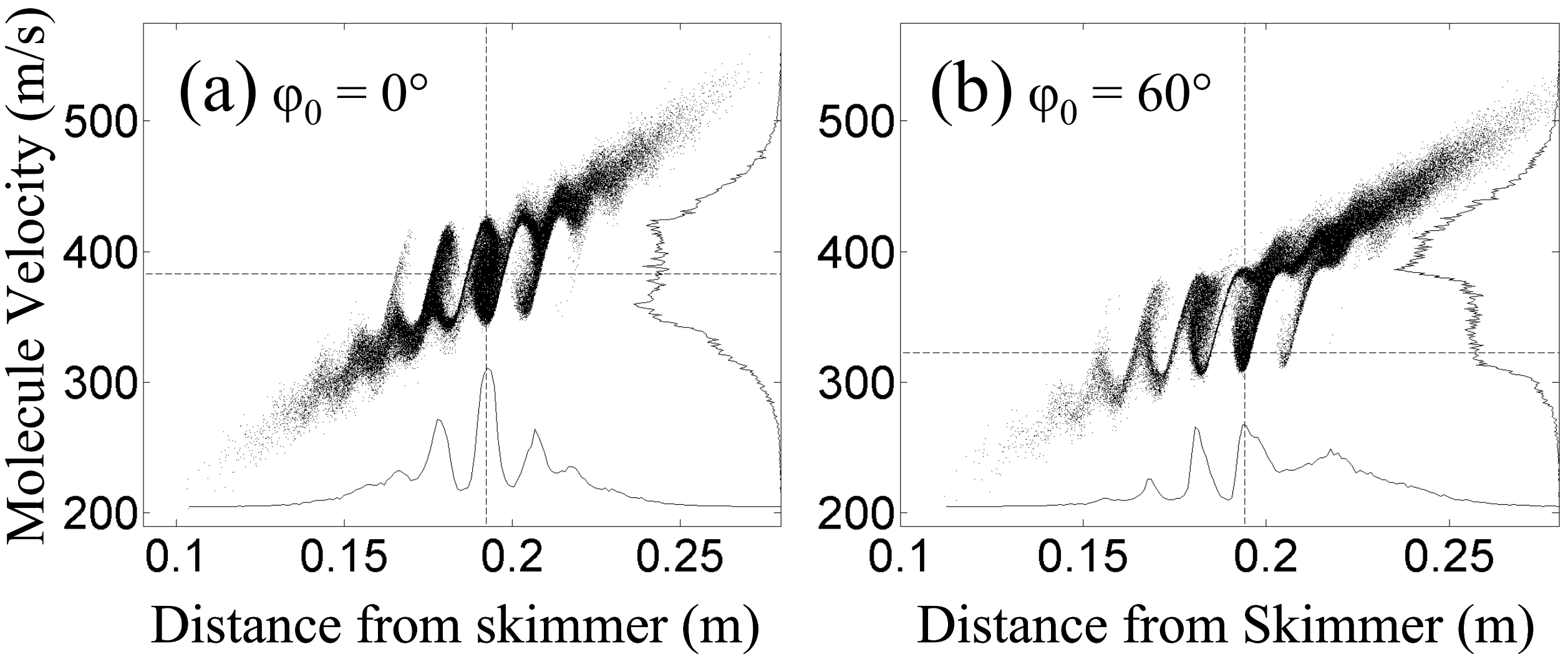

Figure 12: 3-D Monte Carlo simulation of the OH packet in front of the 23rd stage under the bunching and slowing (60°) conditions. Molecules are shown in longitudinal phase space as dots. Projections along the longitudinal spatial and velocity coordinates are both shown.

Figure 12 Bochinski et al.

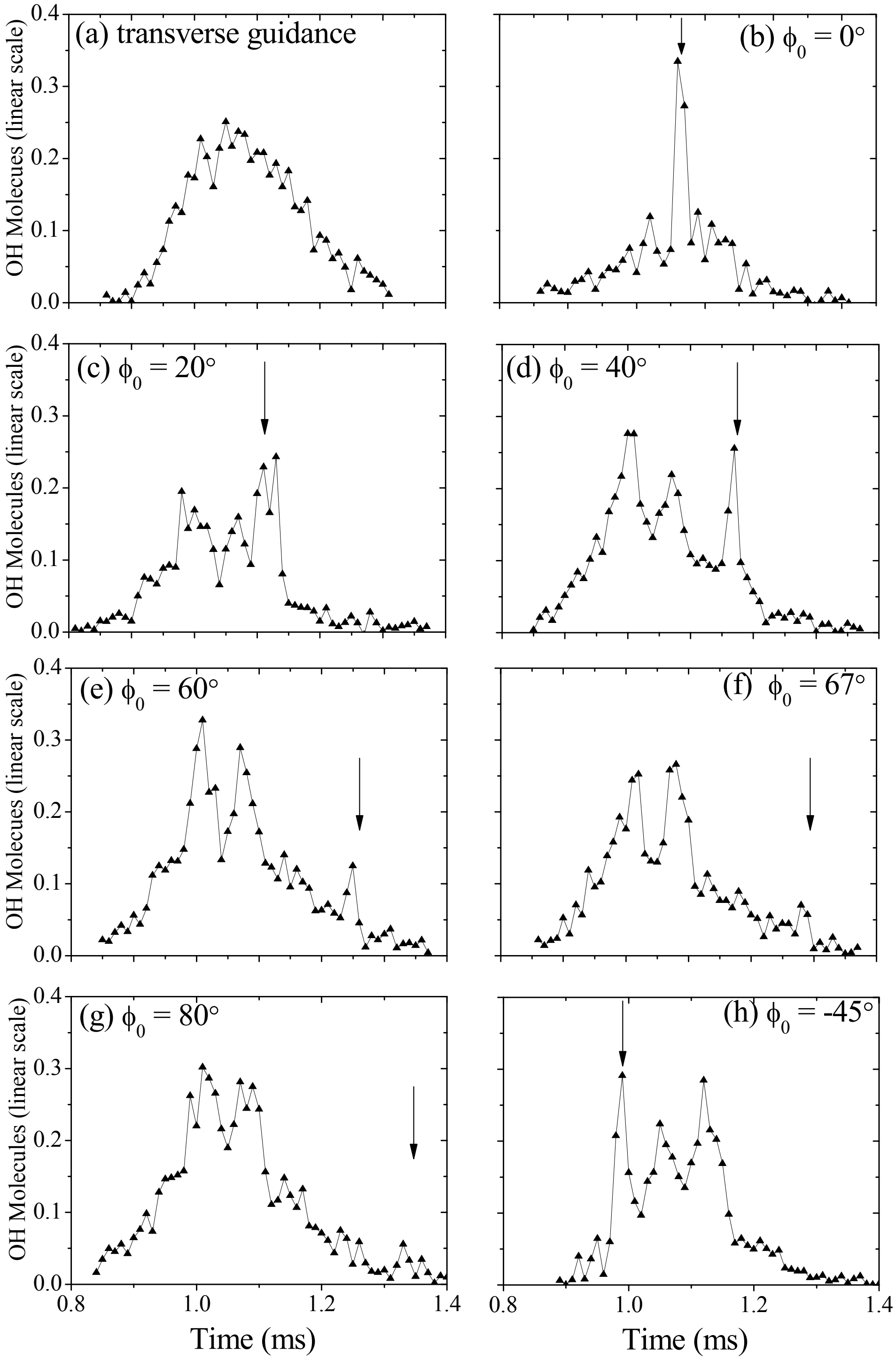

Figure 13: TOF LIF measurements of OH molecules before the 51st stage under (a) transverse guidance, (b) $\varphi_0$ = 0° [385 m/s], (c) $\varphi_0$ = 20° [337 m/s], (d) $\varphi_0$ = 40° [283 m/s], (e) $\varphi_0$ = 60° [220 m/s], (f) $\varphi_0$ = 67°[198 m/s], (g) $\varphi_0$ = 80° [168], and (h) $\varphi_0$ = -45° [470], where the number in square brackets indicates the mean velocity of the phase stable packet.

Figure 13 Bochinski et al.

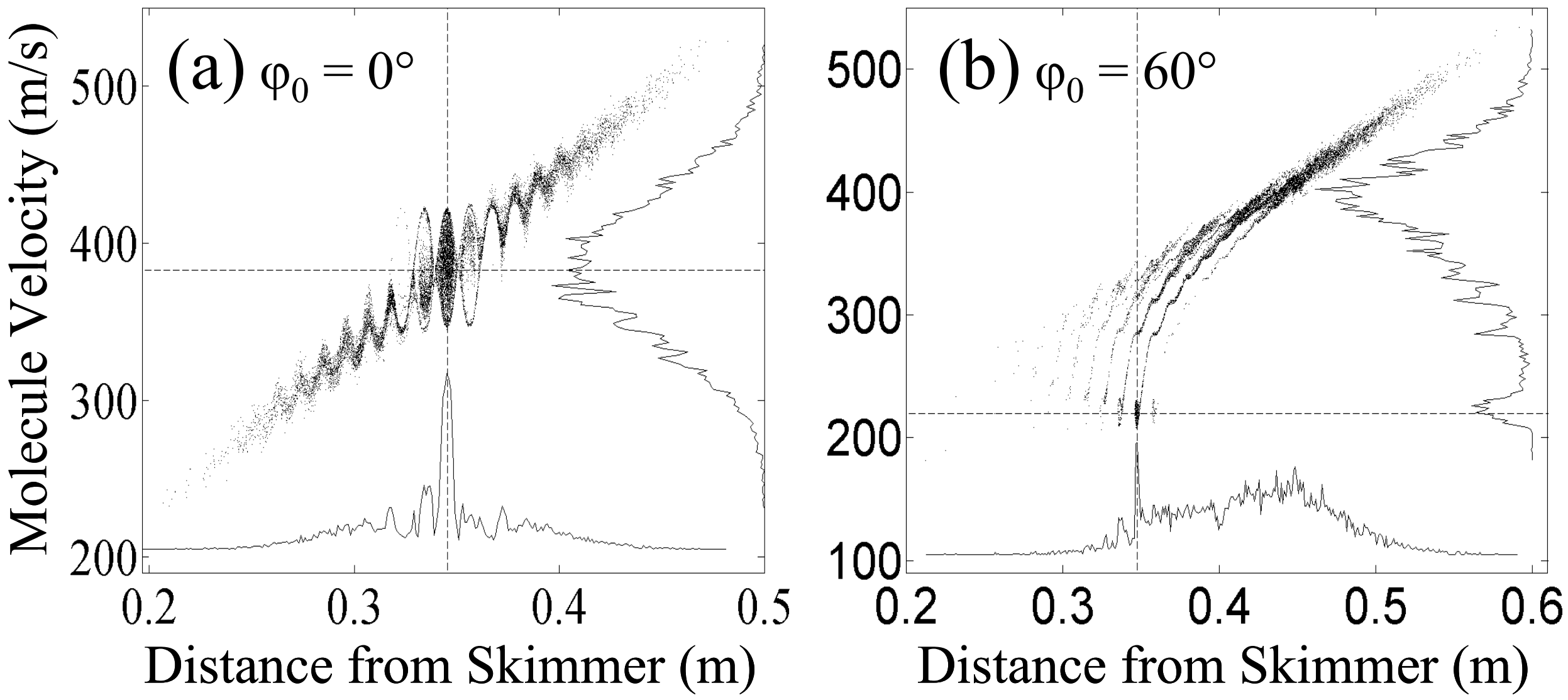

Figure 14: 3-D Monte Carlo simulation of the OH packet in front of the 51st stage under the bunching and slowing (60°) conditions. Molecules are shown in longitudinal phase space as dots. Projections along the longitudinal spatial and velocity coordinates are both shown.

Figure 14 Bochinski et al.

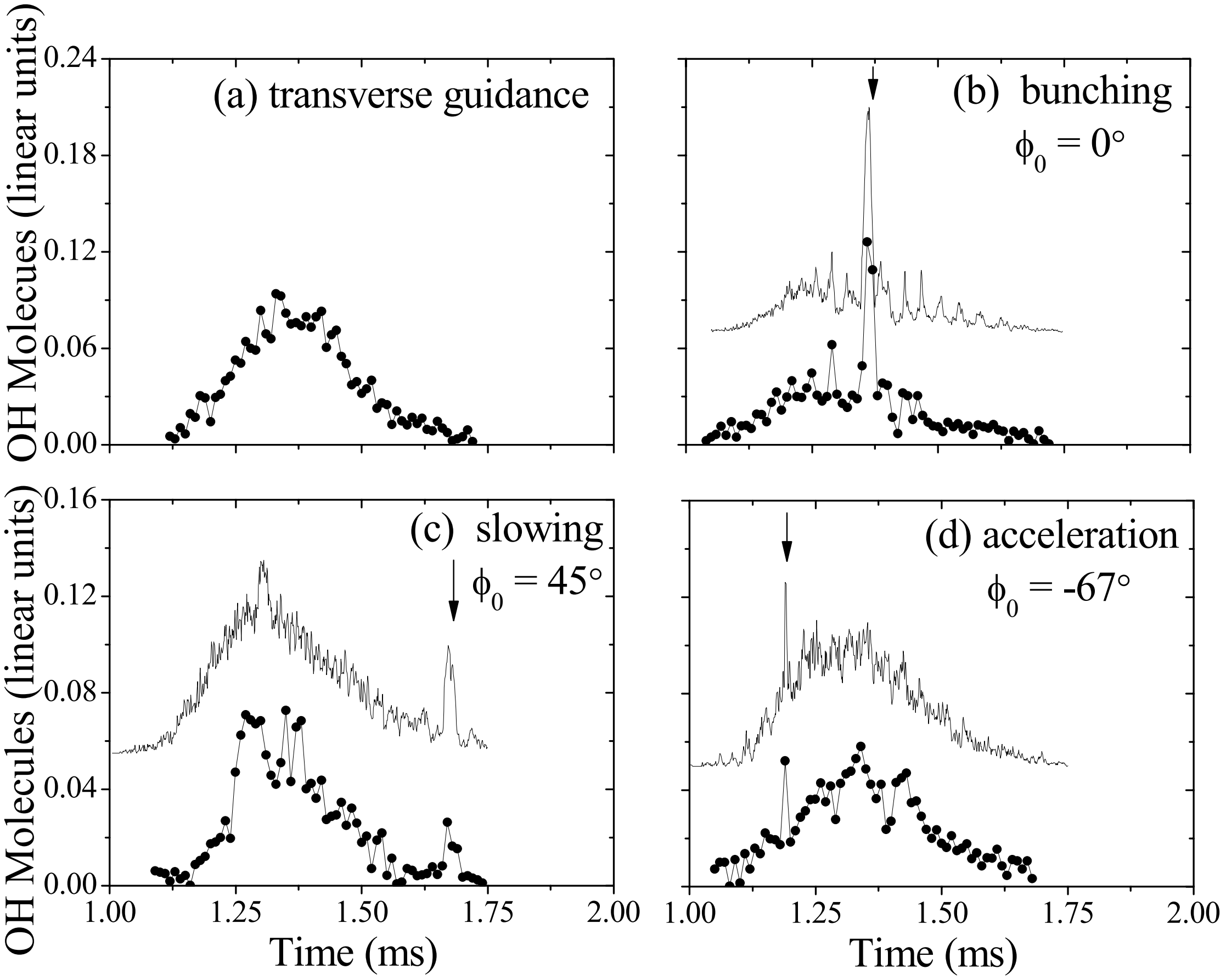

Figure 15: Observations of OH molecules moving through the trap region under (a) transverse guidance, (b) bunching ($\varphi_0$ = 0°, v = 385 m/s), (c) slowing ($\varphi_0$ = 45°, v = 210 m/s), and (d) acceleration ($\varphi_0$ = -67°, v = 541 m/s). Monte Carlo simulation traces are plotted above the data.

Figure 15 Bochinski et al.

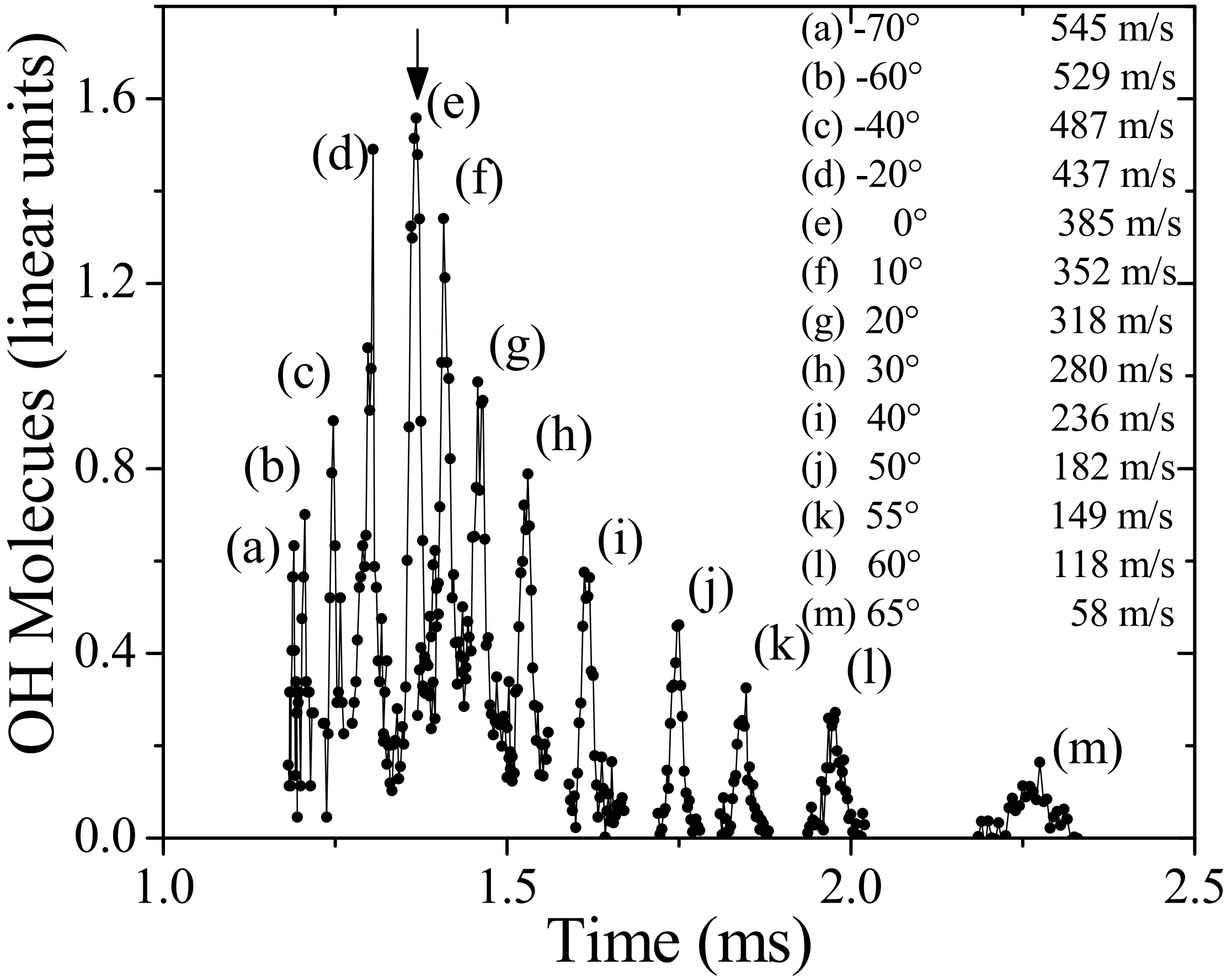

Figure 16: Slowed, accelerated, and bunched molecular packages traversing the trap region at varying phase angles. Here, the downward-pointing arrow indicates solely the position of the bunched (trace (e) $\varphi_0 = 0°$) packet, as all observed peaks are the result of phase-stable packets. Molecular peaks arriving earlier in time result from accelerated OH packets with corresponding negative phase-angles as denoted in the figure. Peaks arriving later in time are similarly the result of slowed molecules, generated by using positive phase-angles.

Figure 16 Bochinski et al.

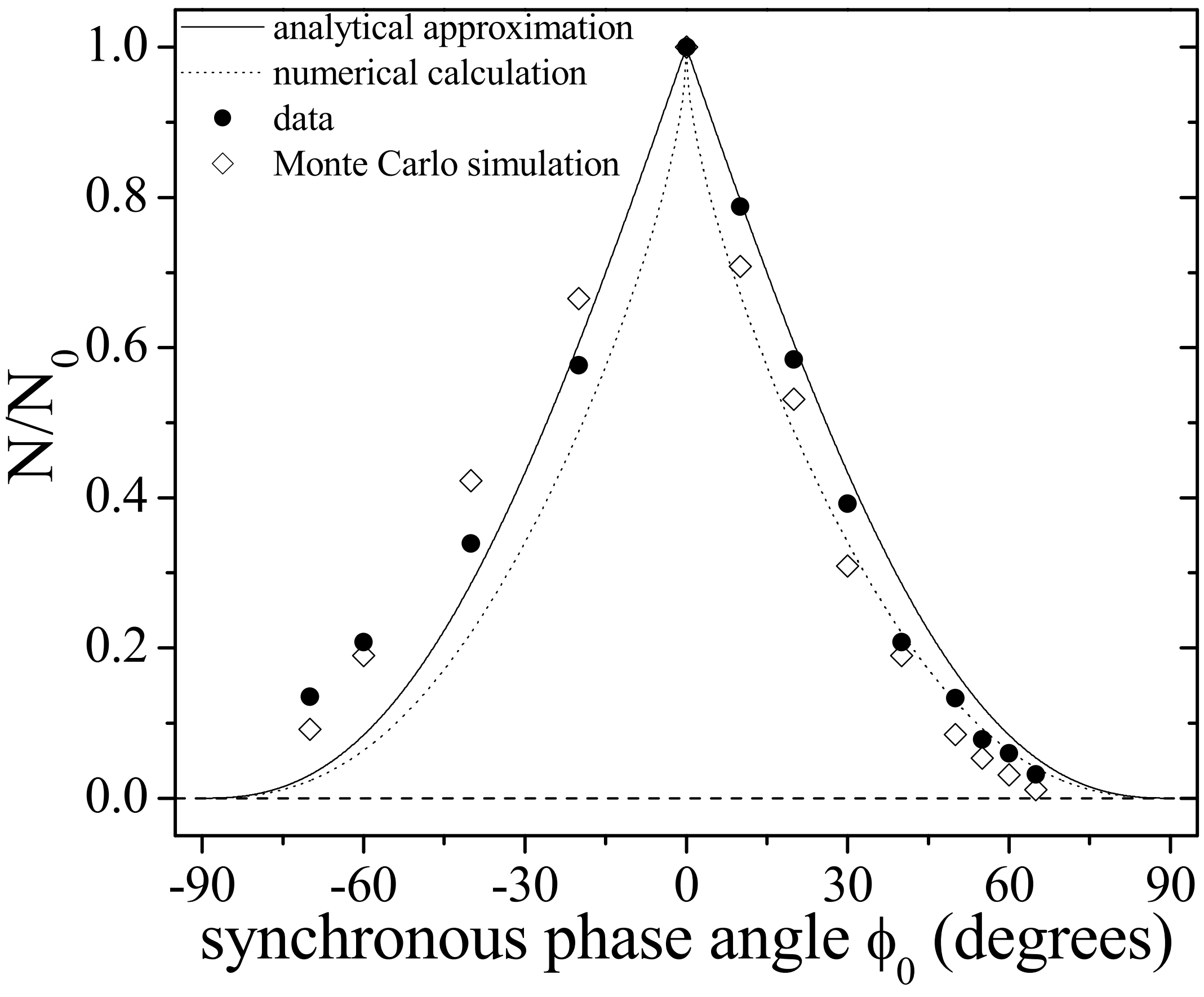

Figure 17: Observations within the trap region: numbers of molecules in the phase stable packet N relative to the number in the bunched packet $N_0$, for varying slowing and accelerating phase angles. Data points (solid circles) and three-dimensional Monte Carlo simulations (open diamonds) are shown. For comparison, theory traces generated from the simple analytic expression (Eq. 8, solid line) and the numerical calculation (dotted line) for stable areas in phase space are shown.

Figure 17 Bochinski et al.